%% file: GB_paper.tex
\documentclass[10pt,conference]{IEEEtran}
\IEEEoverridecommandlockouts

\input{preamble.tex}

\begin{document}

\title{
Real-time Bayesian inference at extreme scale:\\ 
A digital twin for tsunami early warning
applied to the Cascadia subduction zone
\thanks{\hyphenpenalty=10000
This research was supported by DOD MURI grants FA9550-21-1-0084 and FA9550-24-1-0327, DOE ASCR grant DE-SC0023171, NSF grants DGE-2137420, EAR-2225286, EAR-2121568, OAC-2139536, and EU's Horizon 2020 grants 101093038, 101058518, 101058129. 
Work performed under the auspices of the U.S.\ Department of Energy
by Lawrence Livermore National Laboratory under Contract DE-AC52-07NA27344 (LLNL-PROC-2004753). 
This research used resources from the Swiss National Supercomputing Centre (CSCS), the Texas~Advanced~Computing~Center (TACC), the National Energy Research Scientific Computing Center (NERSC, ALCC-ERCAP0030671), and Livermore~Computing (LC).
}
}

\author{
\IEEEauthorblockN{Stefan Henneking}
\IEEEauthorblockA{
\textit{Oden Institute}\\
\textit{The University of Texas at Austin}\\
stefan@oden.utexas.edu}
\and
\IEEEauthorblockN{Sreeram Venkat}
\IEEEauthorblockA{
\textit{Oden Institute}\\
\textit{The University of Texas at Austin}\\
srvenkat@utexas.edu}
\and
\IEEEauthorblockN{Veselin Dobrev}
\IEEEauthorblockA{\textit{Center for Applied Scientific Computing} \\
\textit{Lawrence Livermore National Laboratory}\\
dobrev1@llnl.gov}
\and
\IEEEauthorblockN{John Camier}
\IEEEauthorblockA{\textit{Center for Applied Scientific Computing} \\
\textit{Lawrence Livermore National Laboratory}\\
camier1@llnl.gov}
\and
\IEEEauthorblockN{Tzanio Kolev}
\IEEEauthorblockA{\textit{Center for Applied Scientific Computing} \\
\textit{Lawrence Livermore National Laboratory}\\
kolev1@llnl.gov}
\and
\IEEEauthorblockN{Milinda Fernando}
\IEEEauthorblockA{
\textit{Oden Institute}\\
\textit{The University of Texas at Austin}\\
milinda@oden.utexas.edu}
\and
\IEEEauthorblockN{Alice-Agnes Gabriel}
\IEEEauthorblockA{\textit{Scripps Institution of Oceanography} \\
\textit{University of California San Diego}\\
algabriel@ucsd.edu}
\and
\IEEEauthorblockN{Omar Ghattas}
\IEEEauthorblockA{
\textit{Oden Institute, Walker Department of Mechanical Engineering}\\
\textit{The University of Texas at Austin}\\
omar@oden.utexas.edu}
}

\maketitle


\begin{abstract}
We present a Bayesian inversion-based digital twin that
employs acoustic pressure data from seafloor sensors, along with 3D
coupled acoustic--gravity wave equations, to infer earthquake-induced
spatiotemporal seafloor motion in real time and forecast tsunami
propagation toward coastlines for early warning with quantified
uncertainties. Our target is the Cascadia subduction zone, with one
billion parameters. Computing the posterior mean alone would require
50 years on a 512~GPU machine. Instead, exploiting the
shift invariance of the parameter-to-observable map and devising novel
parallel algorithms, we induce a fast offline--online
decomposition. The offline component requires just one adjoint wave
propagation per sensor; using MFEM, we scale this part of the computation
to the full El Capitan system (43,520 GPUs) with 92\% weak
parallel efficiency. Moreover, given real-time data, the online
component exactly solves the Bayesian inverse and forecasting problems
in 0.2~seconds on a modest GPU system, a ten-billion-fold speedup.
\end{abstract}

\vspace*{-5pt}

\vspace*{-5pt}
\section{Justification for ACM Gordon Bell Prize}
\vspace*{-1pt}
Fastest time-to-solution of a PDE-based Bayesian inverse problem
with 1 billion parameters in 0.2~seconds, a
ten-billion-fold speedup over SoA.
Largest-to-date unstructured mesh FE simulation with 
55.5 trillion DOF on 43,520 GPUs, 
with 92\% weak and 79\% strong parallel efficiencies 
in scaling over a 
128$\times$ increase of GPUs
on the full-scale \emph{El Capitan} system.




\section{Performance Attributes}

\vspace{-8pt}
\begin{table}[htb]
\centering
\begin{tabular}{@{}ll@{}}
    \toprule
    Performance attribute & This submission \\
    \midrule
    Category of achievement & Scalability, time-to-solution, peak performance \\
    Type of method used & Bayesian inversion, real-time computing, FEM \\
    Results reported based on & Whole application including I/O \\
    Precision reported & Double precision \\
    System scale & Results measured on full-scale system \\
    Measurement mechanism & Timers, DOF throughput, FLOP count \\
    \bottomrule
\end{tabular}
\end{table}
\vspace{-4pt}


\section{Overview of the Problem}
\label{sec:overview}
\input{3_overview.tex}


\section{Current State of the Art}
\label{sec:state_of_the_art}
\input{4_state_of_the_art.tex}

\section{Innovations Realized}
\label{sec:innovations}
\input{5_innovations.tex}

\section{How Performance Was Measured}
\label{sec:measurement}
\input{6_measurement.tex}

\section{Performance Results}
\label{sec:results}
\input{7_results.tex}


\section{Implications}
\label{sec:implications}
\input{8_implications.tex}

\section*{Acknowledgment}
The authors thank Pythagoras Watson and Livermore Computing for assistance with \emph{El Capitan} runs, Daniel Ganellari and Sebastian Keller for assistance with \emph{Alps} runs, Amanda Dufek and Muaaz Awan for assistance with \emph{Perlmutter} runs, John Cazes for assistance with \emph{Frontera} runs, Jonatan Glehman for assistance in generating earthquake simulation output, and Eric Dunham for insightful discussions about this work.



\printbibliography

\end{document}

%% file: preamble.tex
\usepackage{amsmath,amssymb,amsfonts}
\usepackage{algorithmic}
\usepackage{graphicx}
\usepackage{textcomp}
\usepackage[table]{xcolor}
\usepackage{physics}
\usepackage{url}
\usepackage{tabularray}
\usepackage{multirow}
\def\BibTeX{{\rm B\kern-.05em{\sc i\kern-.025em b}\kern-.08em
    T\kern-.1667em\lower.7ex\hbox{E}\kern-.125emX}}

\usepackage[ruled]{algorithm2e}
\usepackage{booktabs}
\usepackage{mathtools}
\usepackage{comment}

\usepackage[compact]{titlesec}
\titlespacing*{\section}{0pt}{1.6ex plus 0.4ex minus 0.2ex}{1.0ex plus .2ex minus .2ex}
\titlespacing*{\subsection}{0pt}{1.2ex plus 0.6ex minus 0.1ex}{.9ex plus .3ex minus .1ex}

\usepackage[backend=biber,
    sorting=none,
    maxbibnames=5,
    maxcitenames=5,
    giveninits=true,
    isbn=false,
    url=false,
    doi=false]
    {biblatex}
\usepackage[hidelinks]{hyperref}
\hypersetup{colorlinks=true, citecolor=blue, linkcolor=blue, urlcolor=blue}
\usepackage{cleveref}

\definecolor{RED}{RGB}{190, 30, 49}
\definecolor{BLUE}{RGB}{11, 78, 179}
\definecolor{GREEN}{RGB}{0, 171, 86}
\definecolor{OLIVE}{RGB}{93, 150, 72}
\definecolor{GREY}{RGB}{50, 50, 50}
\definecolor{BROWN}{RGB}{122, 90, 40}

\DeclareMathOperator*{\argmin}{arg\,min}
\def\lb{\langle}
\def\rb{\rangle}

\newcommand{\bb}[1]{\mathbb{#1}}
\newcommand{\mc}[1]{\mathcal{#1}}

\newcommand{\tc}[1]{\textcolor{#1}}

\newcommand{\be}{\begin{equation}}
\newcommand{\ee}{\end{equation}}

\newcommand{\bes}{\begin{equation*}}
\newcommand{\ees}{\end{equation*}}

\newcommand{\eq}[1]{(\ref{eq:#1})}
\newcommand{\fig}[1]{\ref{fig:#1}}
\newcommand{\tab}[1]{\ref{tab:#1}}

\newcommand{\p}{\partial}

\newcommand{\paramtxt}{m}
\newcommand{\datatxt}{d}
\newcommand{\paramvec}{\vb{\paramtxt}}
\newcommand{\datavec}{\vb{\datatxt}}
\newcommand{\obsvec}{\vb{\datatxt}^{\text{obs}}}

\newcommand{\qoivec}{\vb{q}}

\newcommand{\Kmat}{\mathbf{K}}
\newcommand{\Qmat}{\mathbf{Q}}
\newcommand{\ptomat}{\mathbf{F}}
\newcommand{\ptqmat}{\mathbf{F}_{\hskip -1pt q}}
\newcommand{\Gmat}{\mathbf{G}}

\newcommand{\numtime}{N_t}
\newcommand{\numparam}{N_m}
\newcommand{\numdata}{N_d}
\newcommand{\numqoi}{N_q}

\newcommand{\noisevec}{\boldsymbol{\nu}}

\newcommand{\bfnoise}{\mathbf{\Gamma}_{\hskip -2pt \text{noise}}}
\newcommand{\bfpriorcov}{\mathbf{\Gamma}_{\hskip -1pt \text{prior}}}

\newcommand{\piprior}{\pi_{\text{prior}}}
\newcommand{\pipost}{\pi_{\text{post}}}
\newcommand{\pilike}{\pi_{\text{like}}}

\newcommand{\bfmmap}{\vb{m}_{\text{map}}}
\newcommand{\bfmprior}{\vb{m}_{\text{prior}}}

\newcommand{\bfpostcovm}{\mathbf{\Gamma}_{\hskip -2pt \text{post}}}
\newcommand{\bfpriorcovm}{\mathbf{\Gamma}_{\hskip -2pt \text{prior}}}

\newcommand{\bfpostcovq}{\mathbf{\Gamma}_{\hskip -2pt \text{post}(q)}}

\newcommand{\bfqmap}{\vb{q}_{\text{map}}}

%% file: 3_overview.tex

\subsection{Tsunami forecasting and the Cascadia subduction zone}

\begin{figure*}[htb]
    \centering
    \begin{minipage}[b]{0.481\textwidth}
    \includegraphics
    [width=\textwidth,trim={0 10pt 0 0}]
    {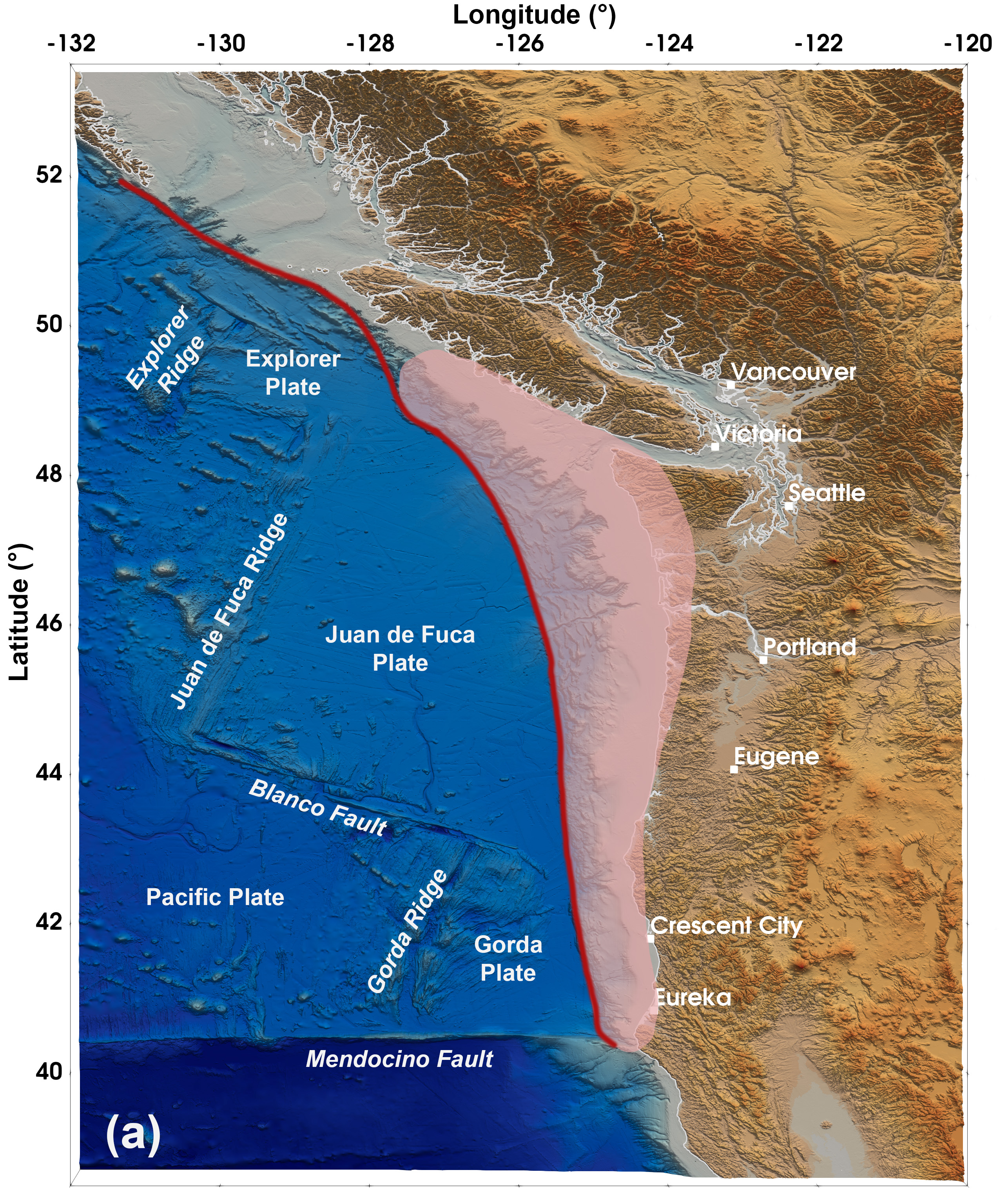}%
    \end{minipage}%
    \hfill
    \begin{minipage}[b]{0.512\textwidth}
    \includegraphics
    [width=0.49\textwidth,trim={0 70pt 0 0},clip]
    {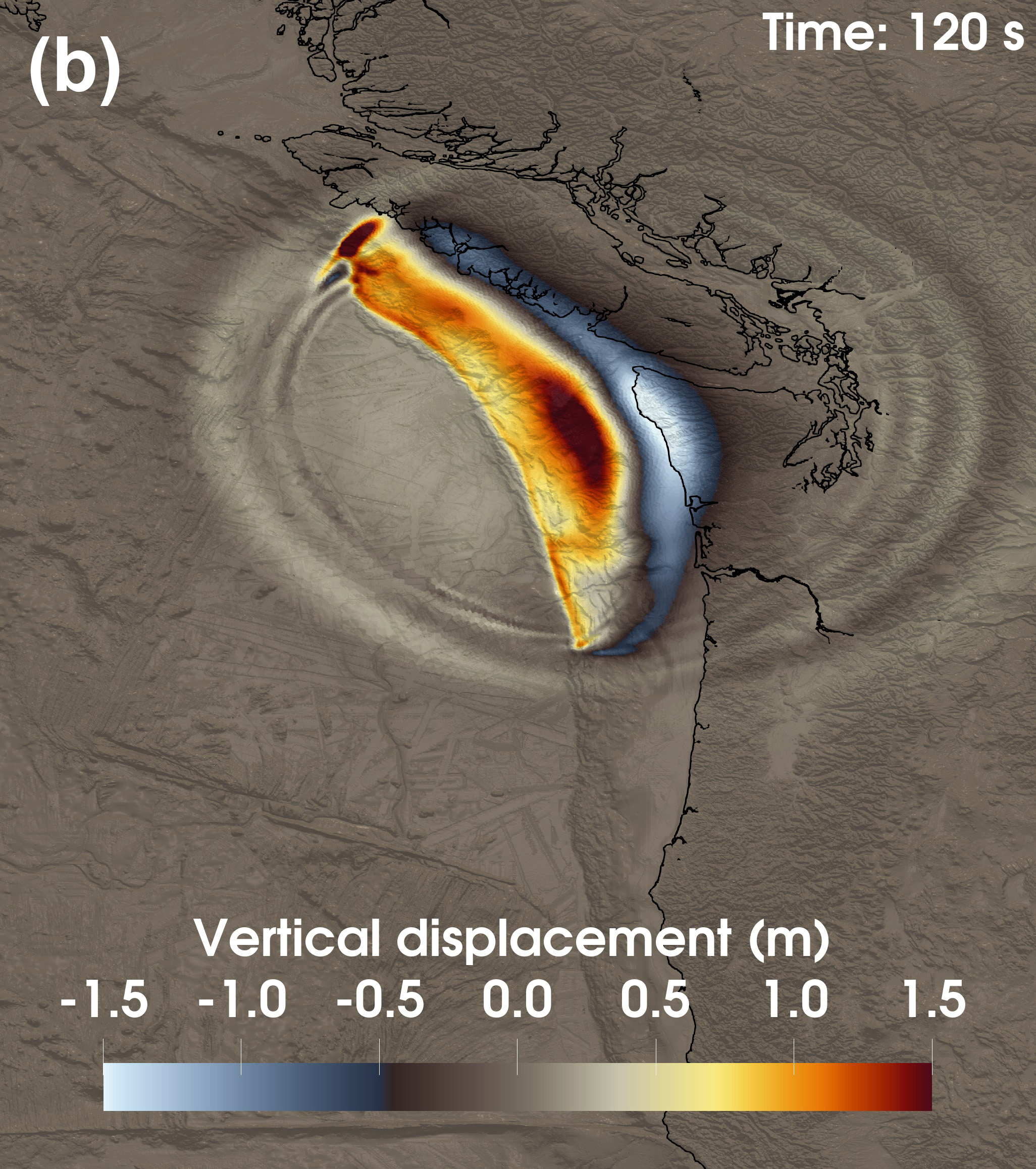}%
    \hfill
    \includegraphics
    [width=0.49\textwidth,trim={0 70pt 0 0},clip]
    {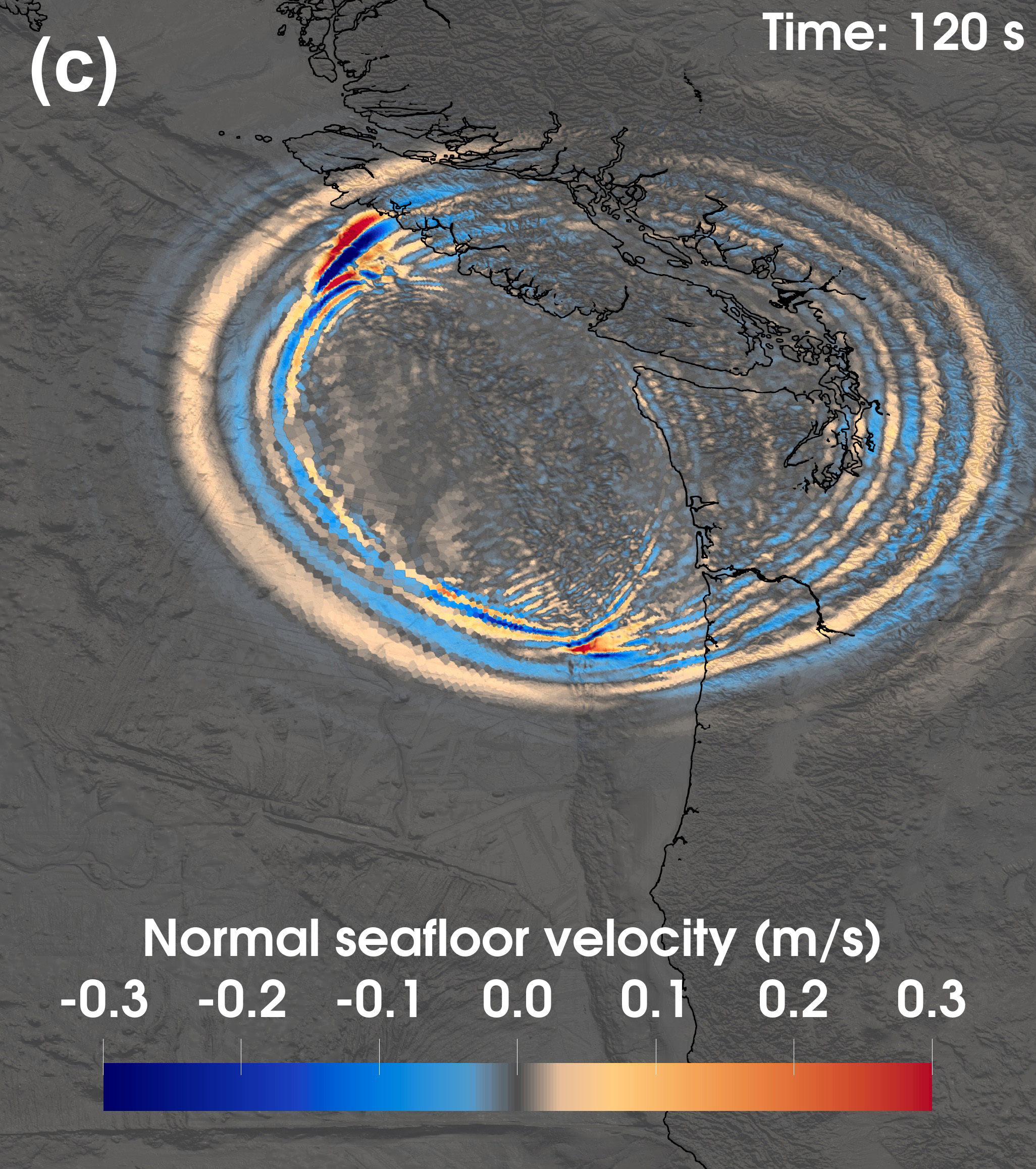}
    \includegraphics
    [width=\textwidth,trim={50pt 180pt 0pt -40pt},clip]
    {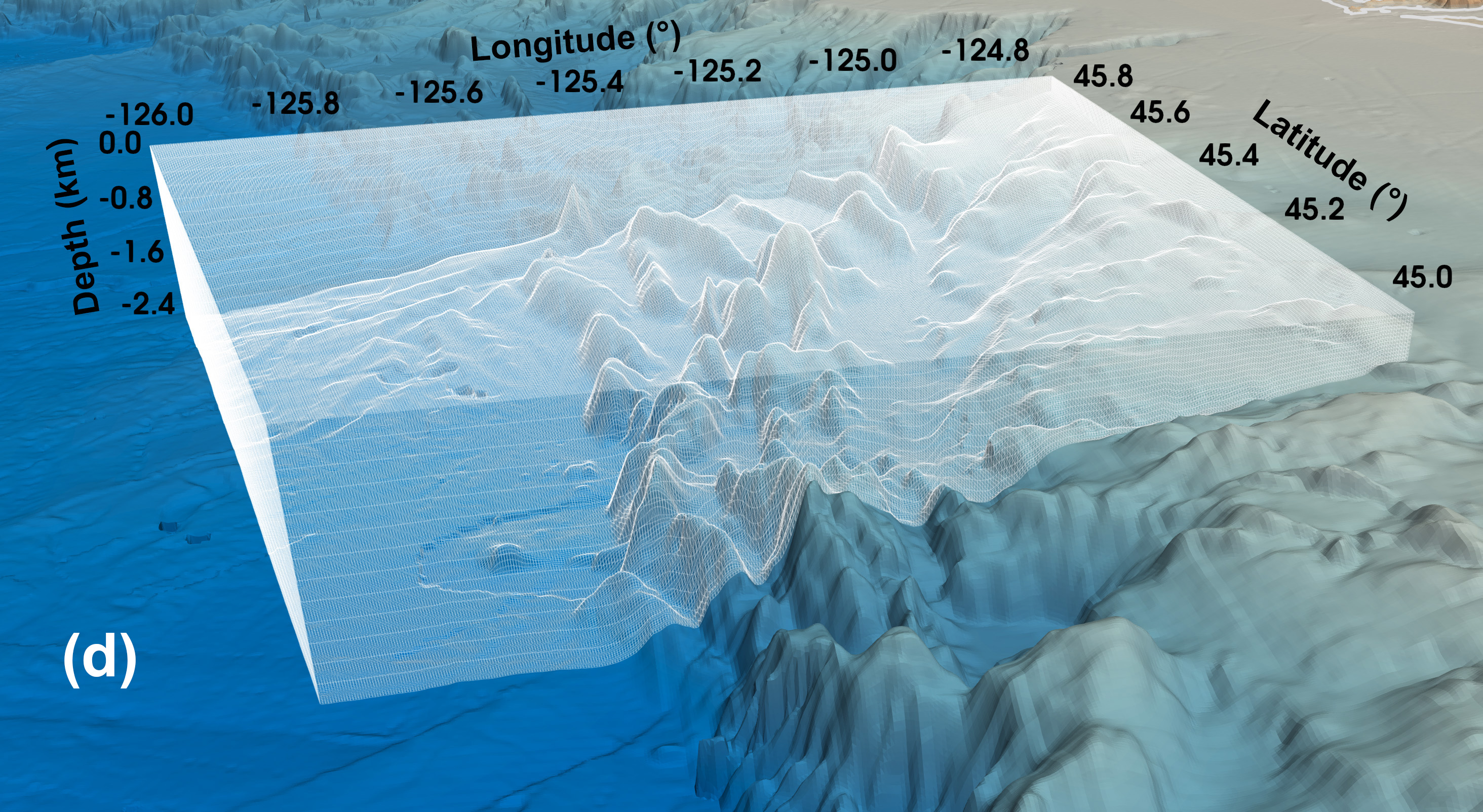}%
    \end{minipage}%
    \caption{
    (a) Topobathymetric map of the Cascadia subduction zone (CSZ) based on GEBCO's (General Bathymetric Chart of the Oceans) 15-arc-second gridded bathymetric data set. The shaded area illustrates the potentially ``locked'' portion of the megathrust interface. The red line marks the trench where the subducting 
    Explorer, Juan de Fuca, and Gorda plates 
    begin their descent beneath the North American Plate.
    (b)--(c)
    Snapshots of vertical seafloor uplift and seafloor velocity from a physics-based 3D dynamic rupture and seismic wave propagation computation of a magnitude 8.7 earthquake scenario spanning the full margin of the CSZ.
    (d)
    Representative block of a 3D multi-block hexahedral mesh of the CSZ, depicting bathymetry-adapted meshing.
    }
    \label{fig:csz-bathymetry}
\end{figure*}

Tsunamis are rare events but can be extremely deadly and cause catastrophic socioeconomic losses. In the past century tsunamis have claimed more than 400,000 lives, and individual tsunamis can result in hundreds of billions of dollars in damage, making them among the costliest natural disasters.
State-of-the-art (SoA) tsunami early warning systems \cite{Kamigaichi2009} rely on rapid characterization of source parameters primarily from seismic data \cite{Hirshorn2021}, issuing alerts within minutes but based on simplified assumptions about the event. These systems typically infer earthquake moment magnitude and hypocenter location, and then trigger tsunami forecasts using precomputed scenarios or simplified fault models, assuming instantaneous, uniform slip on a predefined fault geometry.
This approach is inadequate in the near field \cite{leveque2018cascadia}, where destructive tsunami waves can arrive on-shore in under ten minutes.
In addition, current models fail to capture the complexities of earthquake rupture dynamics \cite{uphoff2017sumatra}, including variable slip distributions or slow rupture speeds. These factors can lead to delayed, missed, or false warnings, especially when early seismic data misrepresent the true tsunami potential, such as during the 2011 Tōhoku, Japan tsunami or the 2024 Cape Mendocino, California earthquake, which did not cause a tsunami, despite five million people receiving evacuation alerts. 
To improve warning capabilities, real-time ocean bottom pressure data and GPS measurements can be integrated directly to constrain tsunami sources, without relying on assumed fault geometry or magnitude \cite{Tsushima2009,Crowell2024}. 

SoA tsunami simulations typically solve the nonlinear shallow water equations, sometimes with added dispersive terms or empirical corrections, to model wave generation, propagation, and run-up. These models are widely used in tsunami hazard assessment and early warning but are known to omit important physical processes such as wave dispersion, 3D hydrodynamic interactions with bathymetry and coastal infrastructure, and the dynamics of the coupled Earth--ocean system \cite{abrahams2023comparison}. Large-scale tsunami simulations using high-resolution bathymetry are constrained by simplifications in source modeling and may require minutes to hours per run, limiting their use in real-time settings \cite{Behrens2021}.


We present a full-physics Bayesian inversion framework---a so-called {\em digital twin} \cite{DigitalTwinsNASEM}---that enables real-time, data-driven tsunami forecasting with dynamic adaptivity to complex source behavior. 
%
%
%
Our inversion framework employs ocean bottom
pressure data, along with the 3D coupled acoustic–gravity wave equations, to infer 
earthquake-induced spatiotemporal seafloor motion in real time. The solution of this Bayesian inverse problem then provides the boundary forcing to forward propagate the tsunamis toward coastlines and issue forecasts of wave heights at specified locations with quantified uncertainties.
We apply our framework to the Cascadia subduction zone (CSZ), which stretches 1000~km from northern California to British Columbia (Fig.~\ref{fig:csz-bathymetry}). The CSZ has been eerily quiet for over 300 years but is considered overdue for a magnitude 8.0--9.0 megathrust earthquake, with paleoseismic and geodetic evidence indicating a recurrence interval of roughly 250 to 500 years \cite{Atwater1995,Goldfinger2012}. In addition to recently proposed future offshore deployments (e.g., by SZ4D (Subduction Zones in four Dimensions)), the NEPTUNE cabled observatory \cite{Barnes2015} has provided continuous offshore observations since 2009 across the northern Cascadia margin, recording several tsunamis and offering valuable data to inform optimal sensor placement.

\subsection{Bayesian inverse problems}

Our work addresses a critically important class of problems that has received little
attention in the Gordon Bell Prize competition: {\em inverse problems} \cite{Vogel02, ghattas2021learning}, 
in particular in the Bayesian framework
\cite{stuart2010inverse,KaipioSomersalo05}. In a {\em forward problem}, the input {\em parameters} (e.g., initial or boundary  conditions, coefficients, source terms) are specified; solution of the governing equations, here assumed to be 
partial differential equations (PDEs),
then yields the state
variables describing the behavior of the system. However, forward models typically contain uncertain parameters.
Often these uncertain
parameters are infinite-dimensional fields, i.e., they vary in space
(and possibly time). In inverse problems, we seek to infer uncertain parameter fields from measurements or observations of system states.

Infinite-dimensional inverse problems are usually ill-posed: many parameter fields are consistent with
the observational data to within the noise. The classical output least-squares
approach to the inverse problem 
posits a regularization functional
that penalizes unwanted features of the parameter field (e.g.,
magnitude or roughness), yielding a unique 
solution \cite{Vogel02}. In contrast, the {\em Bayesian approach} seeks to
characterize the probability that {\em any} parameter
field is consistent with the data and prior knowledge. 
The uncertain parameter is treated as a random field, and the solution of the inverse problem becomes a probability density,
the
{\em posterior 
distribution} \cite{KaipioSomersalo05}. 
As such the Bayesian approach is very powerful. 

This power comes at a price, though: computing the Bayesian solution is generally intractable for infinite-dimensional inverse
problems governed by PDEs. For example, computing just the mean of the posterior
formally requires numerical integration in the discretized parameter dimension
(in our case, one billion); {\em each evaluation} of the integrand requires solution of the forward problem, which can take hours or more, 
and numerous forward solutions are required. Accordingly, various approximations of the posterior,
such as the Laplace approximation \cite{ghattas2021learning}, must be invoked, and even then
massive-scale computation is necessary, e.g.~\cite{isaac2015scalable,
  Bui-ThanhGhattasMartinEtAl13, CuiLeiLiuEtAl2025}. The
specific Bayesian inverse problem we address here is the
inference and prediction of tsunamis for early warning,
adding a real-time requirement to the challenges 
described above. Our forecasting goal---to provide early warning in less than a minute---may seem futile, as SoA inversion methods (see \S\ref{sec:state_of_the_art}) would require {\em 50 years} on 512 NVIDIA A100 GPUs to solve this problem. 
As we shall see in \S\ref{sec:innovations}, we achieve real-time performance
through a combination of novel parallel inversion algorithms, advanced PDE discretization libraries, and exploitation of leading-edge GPU supercomputers.

\subsection{Inference of seafloor motion with acoustic--gravity model}
\label{sec:acoustic-gravity}

In this section, we describe our target inverse problem. The forward tsunami dynamics are modeled by the acoustic--gravity wave equations that couple the propagation of ocean acoustic waves with surface gravity waves. 
This PDE model is derived by linearization of the conservation of mass and momentum around hydrostatic pressure in the compressible ocean, and the coupling to the surface gravity wave comes from a modified free surface boundary condition at the sea surface~\cite{lotto2015tsunami}.
The model takes the form of a coupled first-order system in the velocity vector field $\vec{u}(\vec{x},t)$, the pressure field $p(\vec{x},t)$, and the surface gravity wave height $\eta(\vec{x},t)$,
\begin{equation}
        \left\{
        \begin{aligned}
                \rho\, \p_t \vec{u} + \nabla p &= 0, & \Omega \times (0,T), \\
                K^{-1} \p_t p + \nabla \cdot \vec{u} &= 0, & \Omega \times (0,T), \\
                p &= \rho g \eta, & \p \Omega_{\text s} \times (0,T), \\
                \p_t \eta &= \vec{u} \cdot \vec{n}, & \p \Omega_{\text s} \times (0,T), \\
                \vec{u} \cdot \vec{n} &= -\p_t b, & \p \Omega_{\text b} \times (0,T), \\
                \vec{u} \cdot \vec{n} &= Z^{-1} p, & \p \Omega_{\text a} \times (0,T), \\
        \end{aligned}
        \right.\label{eq:fwd-pde}
\end{equation}
with homogeneous initial conditions. Here, $K$ and $\rho$ are the bulk modulus and density of seawater, $Z = \rho c$ is the acoustic wave impedance, $c = \sqrt{K/\rho}$ is the speed of sound in seawater, $g$ is gravitational acceleration, and $\p_t b$ is the seafloor inward normal velocity with $b$ the displacement. The spatial domain is denoted by $\Omega$ with boundaries $\p \Omega_{\text s}$ (sea surface), $\p \Omega_{\text b}$ (sea bottom), and $\p \Omega_{\text a}$ (lateral, absorbing boundaries); $\vec{n}$ is the outward unit normal; and the temporal domain is $(0,T)$.

The uncertain parameter field $m(\vec{x},t)$, to be inferred from the data,
is the spatiotemporal seafloor inward normal velocity, 
\bes
    m(\vec{x},t) \coloneq \p b(\vec{x},t) / \p t, \quad
    (\vec{x},t) \in \p \Omega_{\text b} \times (0,T) .
    \label{eq:parameter-field}
\ees
The observables $d$ are model predictions of data $d^\text{obs}$ from pressure sensors mounted on the seafloor, i.e.~$d = 
p(\vec{x}^{\hskip 1pt d}_j, t^d_i)$, where $\vec{x}^{\hskip 1pt d}_j \in \p \Omega_{\text b}$, $j = 1,\ldots,\numdata$, are the seafloor sensor locations and $t^d_i \in (0,T)$, $i = 1,\ldots,\numtime^d$, are the time 
instants at which observations are made. The {\em parameter-to-observable} (p2o) \emph{map} $\mc F$ takes the seafloor velocity $m$ as input, solves the acoustic--gravity wave equations~\eqref{eq:fwd-pde}, and extracts $d$, the pressures at the sensor locations and observation times. 
The {\em quantities of interest} (QoI) are the surface wave height predictions at locations and times of interest, i.e.~$q = 
\eta(\vec x^{\hskip 1pt q}_j, t^q_i)$ where $\vec x^{\hskip 1pt q}_j \in \p \Omega_{\text s}$, $j = 1,\ldots,\numqoi$, and $t^q_i \in (0,T)$, $i = 1,\ldots,\numtime^q$, are specific locations and time instants pertinent to tsunami early warning. 
The {\em parameter-to-QoI} (p2q) \emph{map} $\mc F_q$ takes the inferred seafloor velocity $m$ and forward predicts the QoI $q$. 

The Bayesian inverse problem can be stated as follows: given pressure recordings $d^\text{obs}$ 
from sensors on the seafloor, infer the spatiotemporal seafloor velocity
$m$ in the subduction zone and its uncertainty. The tsunami posterior prediction problem is: given inferred parameters $m$ and their uncertainty, 
forward predict the surface gravity wave heights $q$ and their uncertainty. 
In \S\ref{sec:state_of_the_art}, we discuss why this problem is intractable with the current SoA.

%% file: 4_state_of_the_art.tex

In this section, we discuss SoA methods for solving the Bayesian inverse problem described in \S\ref{sec:acoustic-gravity}. 
%
%
After discretization in space and time, Bayes' theorem gives 
\bes
	\pipost(\paramvec | \obsvec) \propto \pilike(\obsvec | \paramvec) \piprior(\paramvec) ,
\ees
where $\piprior(\paramvec)$ is the prior probability density of the seafloor velocity parameters $\paramvec$,
$\pilike(\obsvec | \paramvec)$ is the likelihood of the observed seafloor pressure data $\obsvec$
given 
$\paramvec$, and $\pipost(\paramvec | \obsvec)$ is the posterior density reflecting the probability of the parameters conditioned on the data.

Taking a Gaussian prior with mean $\bfmprior$ and covariance $\bfpriorcovm$ (here block diagonal, with each block the inverse of an elliptic PDE operator in space representing a Mat\'{e}rn covariance \cite{stuart2010inverse}), along with centered Gaussian additive noise with noise covariance $\bfnoise$,
the posterior is given by:
\bes
	\pipost(\paramvec  | \obsvec) \! \propto
	\! \exp \! \left\{
		\!- \tfrac{1}{2} \| \ptomat \paramvec \!-\! \obsvec \|_{\bfnoise^{-1}}^2
		\!\!- \!\tfrac{1}{2} \| \paramvec \!-\! \bfmprior \|_{\bfpriorcovm^{-1}}^2
	\! \right\} \! .
\ees
Here $\ptomat$
is the discrete p2o map.

The maximum-a-posteriori (MAP) point $\bfmmap$ maximizes the posterior distribution over admissible parameters $\paramvec$, or equivalently solves the quadratic optimization problem
\bes
	\bfmmap \! \coloneq \!
    \argmin_{\paramvec} \!
	\tfrac{1}{2} \| \ptomat \paramvec - \obsvec \|_{\bfnoise^{-1}}^2 +
	\tfrac{1}{2} \| \paramvec - \bfmprior \|_{\bfpriorcovm^{-1}}^2 .
\ees
The MAP point $\bfmmap$ is thus the solution of the linear system
\be
	\left( \ptomat^* \bfnoise^{-1} \ptomat + \bfpriorcovm^{-1} \right) \bfmmap =
	\ptomat^* \bfnoise^{-1} \obsvec + \bfpriorcovm^{-1} \bfmprior,
	\label{eq:mmap-problem}
\ee
where  $\vb H \coloneq ( \ptomat^* \bfnoise^{-1} \ptomat + \bfpriorcovm^{-1} )$ is the Hessian of the negative log-posterior and $\ptomat^*$ is the adjoint of the p2o map. 

The posterior 
is then a {\em Gaussian centered at the MAP point}, i.e.~$\pipost(\paramvec | \obsvec) \propto \mc N(\bfmmap, \bfpostcovm)$, with posterior covariance
\be
	\bfpostcovm \coloneq \vb H^{-1} = \left( \ptomat^* \bfnoise^{-1} \ptomat + \bfpriorcovm^{-1} \right)^{-1} .
	\label{eq:Hessian}
\ee
%
Furthermore, the posterior predictive of the wave height QoI $\qoivec$, which depend linearly on the parameters via the discrete p2q map $\ptqmat$, can be determined as the Gaussian probability density $\pipost(\qoivec | \obsvec) \propto \mc N \left( \bfqmap, \bfpostcovq \right)$, where $\bfqmap = \ptqmat \bfmmap$ and $\bfpostcovq = \ptqmat \bfpostcovm \ptqmat^*$.

Writing expressions \eqref{eq:mmap-problem} for the posterior mean $\bfmmap$ and~\eqref{eq:Hessian} for the posterior covariance $\bfpostcovm$ is easy;
computing them is a monumental challenge. The coefficient matrix of~\eqref{eq:mmap-problem}, the Hessian $\vb H$, is a dense matrix of dimension of the parameters, 
here $10^9 \times 10^9$. Constructing it efficiently via $\ptomat$ requires as many adjoint wave propagation solutions---here, each requiring $\sim$1 hour on 512 NVIDIA A100 GPUs---as there are spatiotemporal data, here $\sim$250,000; this amounts to 25 years.
Even if $\vb H$ could be constructed and stored (4~exabytes), factoring it would require 10 years on a sustained 1~EFLOP/s machine. SoA methods for large-scale inverse problems employ matrix-free conjugate gradients (CG), preconditioned by the prior covariance (thus involving elliptic PDE solves), and yield convergence in a number of iterations of the order of the number of eigenvalues of the prior-preconditioned Hessian of the negative log likelihood, $\widetilde{\vb H}_\text{like} \coloneq \bfpriorcovm^{1/2} \ptomat^* \bfnoise^{-1} \ptomat \bfpriorcovm^{1/2}$, that are above unity \cite{ghattas2021learning}. For highly ill-posed inverse problems, the eigenvalues 
decay rapidly and $\widetilde{\vb H}_\text{like}$ has low effective rank, and thus CG converges rapidly \cite{ghattas2021learning}. Moreover, a low-rank approximation of $\widetilde{\vb H}_\text{like}$ computed via a matrix-free randomized eigensolver, along with the Sherman--Morrison--Woodbury formula, can be used to compactly represent the posterior covariance, sample from the posterior distribution, and compute the posterior variance field, as has been done for some large-scale Bayesian inverse problems \cite{isaac2015scalable,Bui-ThanhGhattasMartinEtAl13}. 

Unfortunately, our operator is not low rank, due to the hyperbolic nature of the forward wave propagation problem (which tends to preserve information), and the location of the pressure sensors directly on the seafloor, whose motion we seek to infer. In our numerical experiments on smaller but representative regions \cite{henneking2025goal}, the effective rank is nearly of the order of the data dimension, and thus we can expect CG to take $\mc{O}$(250,000) iterations. Since each iteration requires the action of $  \ptomat$ and $\ptomat^*$ on a vector, i.e., a pair of forward/adjoint wave propagations, we would expect CG to take $\mc{O}$(50) years (on 512 GPUs) for our target problem. And this is for computing  $\bfmmap$ via \eqref{eq:mmap-problem}; in the absence of low-rank properties, the  even more daunting task of evaluating the uncertainty of the inverse solution  (which is also needed for the posterior of the QoI, $\bfpostcovq$) via the inverse of the Hessian in \eqref{eq:Hessian} is unthinkable. While approximations of high (global) rank Hessians have been attempted, they have not been viable for high-frequency wave problems in $10^9$ parameter dimensions, e.g., \cite{ambartsumyan2020hierarchical}.


Alternatively, we could relax the need to exactly solve~\eqref{eq:mmap-problem} for $\bfmmap$ and instead seek a surrogate approximation of the p2o map, $\ptomat$. However, this too is a monumental challenge: constructing surrogates in $10^9$ parameter dimensions, where each training point amounts to a forward wave propagation ($\sim$1 hour on 512 GPUs), is essentially impossible using conventional surrogate methods. Specially-architected neural operator surrogates 
that exploit the geometry of p2o maps have been developed, scaled to high ambient dimensions, and deployed within Bayesian inverse problems, e.g.,
\cite{CaoOLearyRoseberryGhattas25}. However, the efficiency of these surrogates (measured by number of training data needed) is predicated on low-rank representations of $\ptomat$, again a property  not enjoyed by our p2o maps. Thus, surrogates are not a viable option for our problem.  

Finally, we might attempt to construct a projection-based reduced order model (ROM) of the forward acoustic--gravity wave equations, by projecting \eqref{eq:fwd-pde} onto a low-dimensional reduced basis. The ROM would have to be parameterized over $10^9$ inputs, which is well beyond the reach of current methods. A more fundamental difficulty is that efficient ROMs for high-frequency wave propagation are not viable due to the Kolmogorov $N$-width problem \cite{greif2019kolmogorov}, which asserts the non-existence of an accurate low-dimensional subspace embedding. 

Next, we present novel parallel algorithms that exploit the autonomous nature of the dynamics \eqref{eq:fwd-pde} and p2o and p2q maps $\ptomat$ and $\ptqmat$ to induce an offline--online decomposition  permitting real-time solution of both the Bayesian inverse problem for seafloor motion and subsequent wave height forecasting.

%% file: 5_innovations.tex
\begin{figure*}[htb]
    \centering
    \includegraphics[width=\textwidth]
    {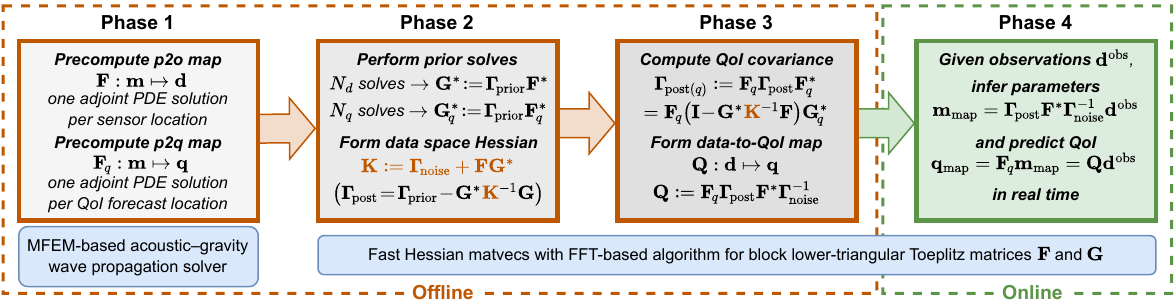}
    \caption{Real-time Bayesian inference framework. The inverse solution is decomposed into several precomputation (offline) Phases 1--3 that are executed just once, and a real-time (online) Phase~4 of parameter inference and QoI prediction that is executed when an earthquake occurs and data are acquired. Phase~1 computes adjoint PDE solutions of the acoustic--gravity model (one PDE solution per sensor and QoI forecast location) to precompute the p2o and p2q block Toeplitz matrices. Phases 2--4 rely on fast FFT-based Hessian actions using this block Toeplitz structure and a transformation of the inverse operator from the high-dimensional parameter space to the  much lower-dimensional data space. Compute times for each phase are given in Table~\ref{tab:time-to-solution} (\S\ref{sec:results}).}
    \label{fig:workflow}
\end{figure*}

Our real-time Bayesian inference and prediction framework combines the following key ideas:
(i)~an offline--online decomposition of the inverse solution that precomputes various mappings to enable the real-time application of the inverse operator and entirely circumvents the need for PDE solutions during the inversion phase; 
(ii)~representation of the inverse operator in data space instead of high-dimensional parameter space;
(iii)~extension of this real-time inference methodology to the goal-oriented setting for the prediction of QoI under uncertainty;
(iv)~exploitation of the shift invariance of 
\eqref{eq:fwd-pde} 
to reduce the required number of offline PDE solutions by several orders of magnitude; and
(v)~exploitation of this same property to perform efficient FFT-based and multi-GPU-accelerated Hessian matrix--vector products (matvecs).
We employ this framework to construct a digital twin for tsunami early warning in the CSZ that solves a Bayesian inverse problem with over $10^9$ parameters and 8,820~QoI \emph{exactly}
(up to rounding errors); 
both the parameter inference and the complete end-to-end, data-to-inference-to-prediction computations can be carried out within a fraction of a second.

\begin{figure*}[htb]
    \includegraphics[width=0.166\textwidth]
    {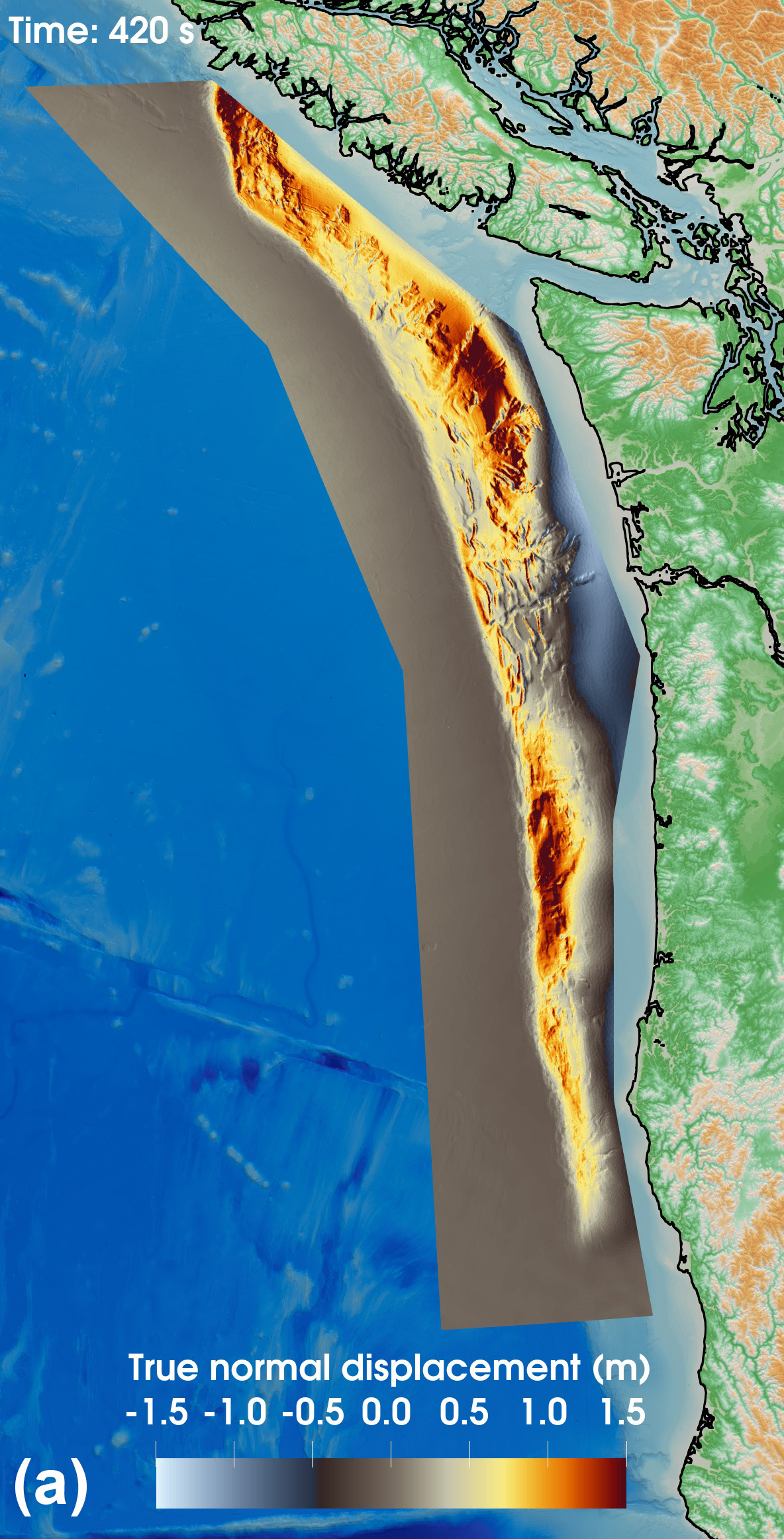}%
    \includegraphics[width=0.166\textwidth]
    {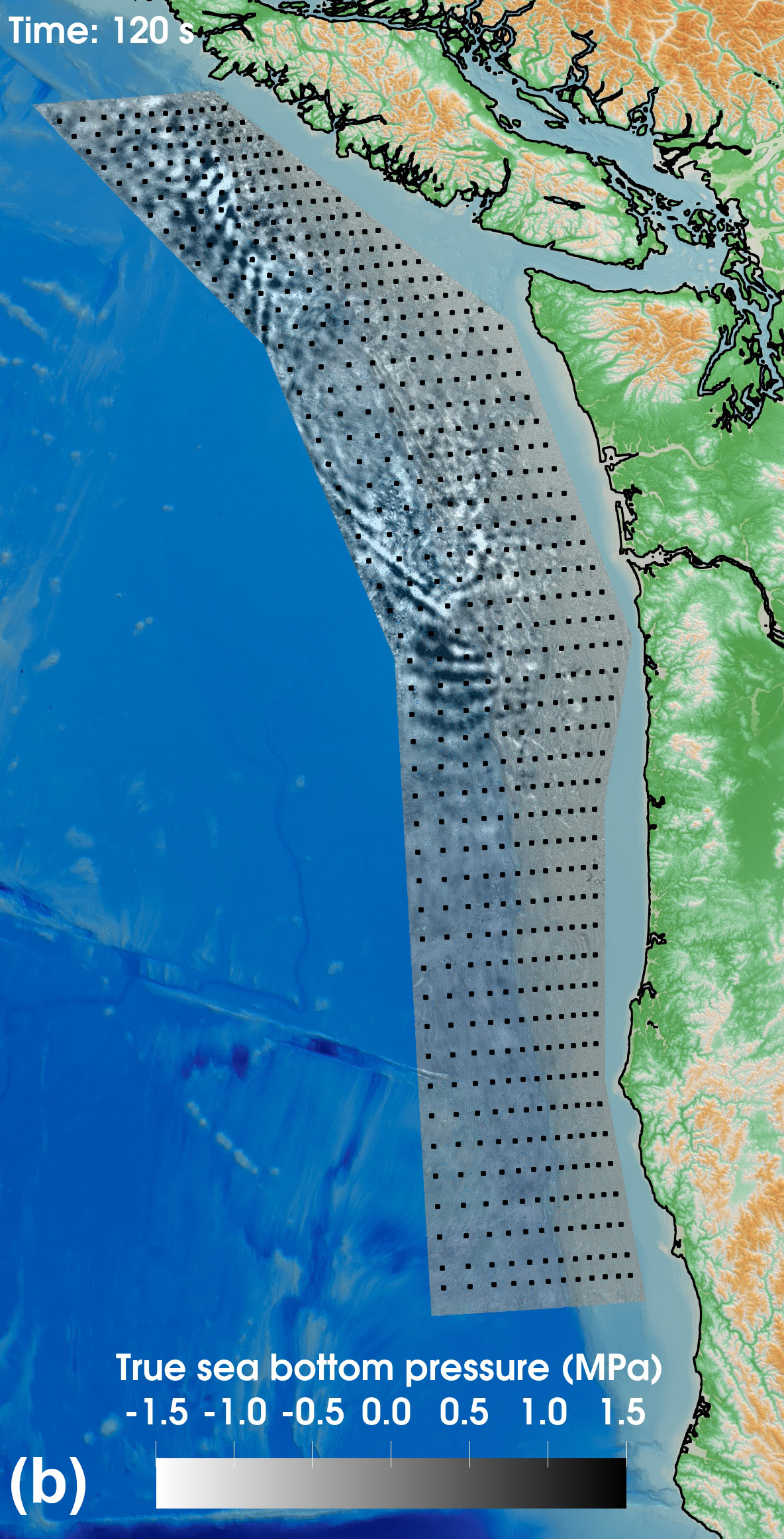}%
    \includegraphics[width=0.166\textwidth]
    {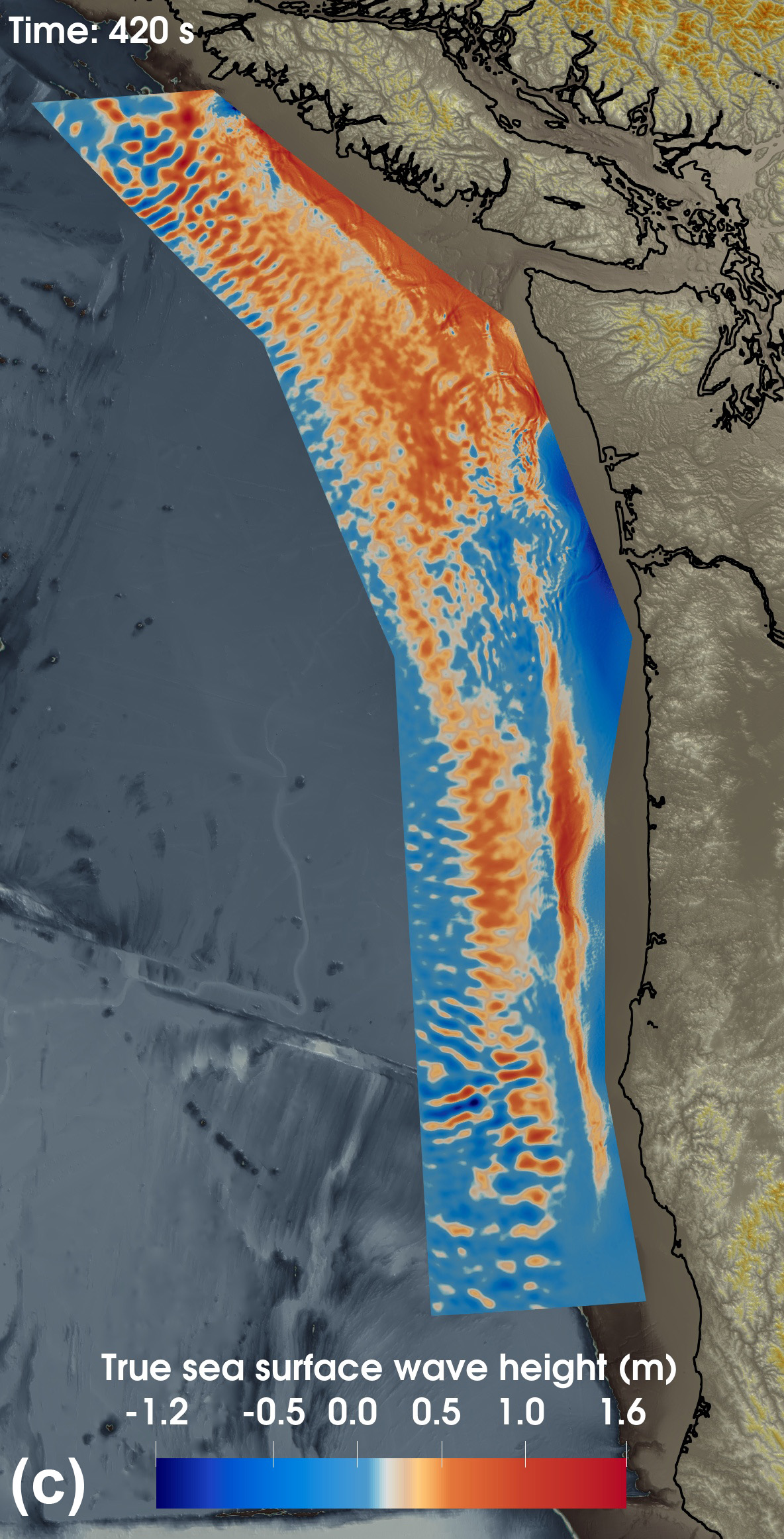}%
    \includegraphics[width=0.166\textwidth]
    {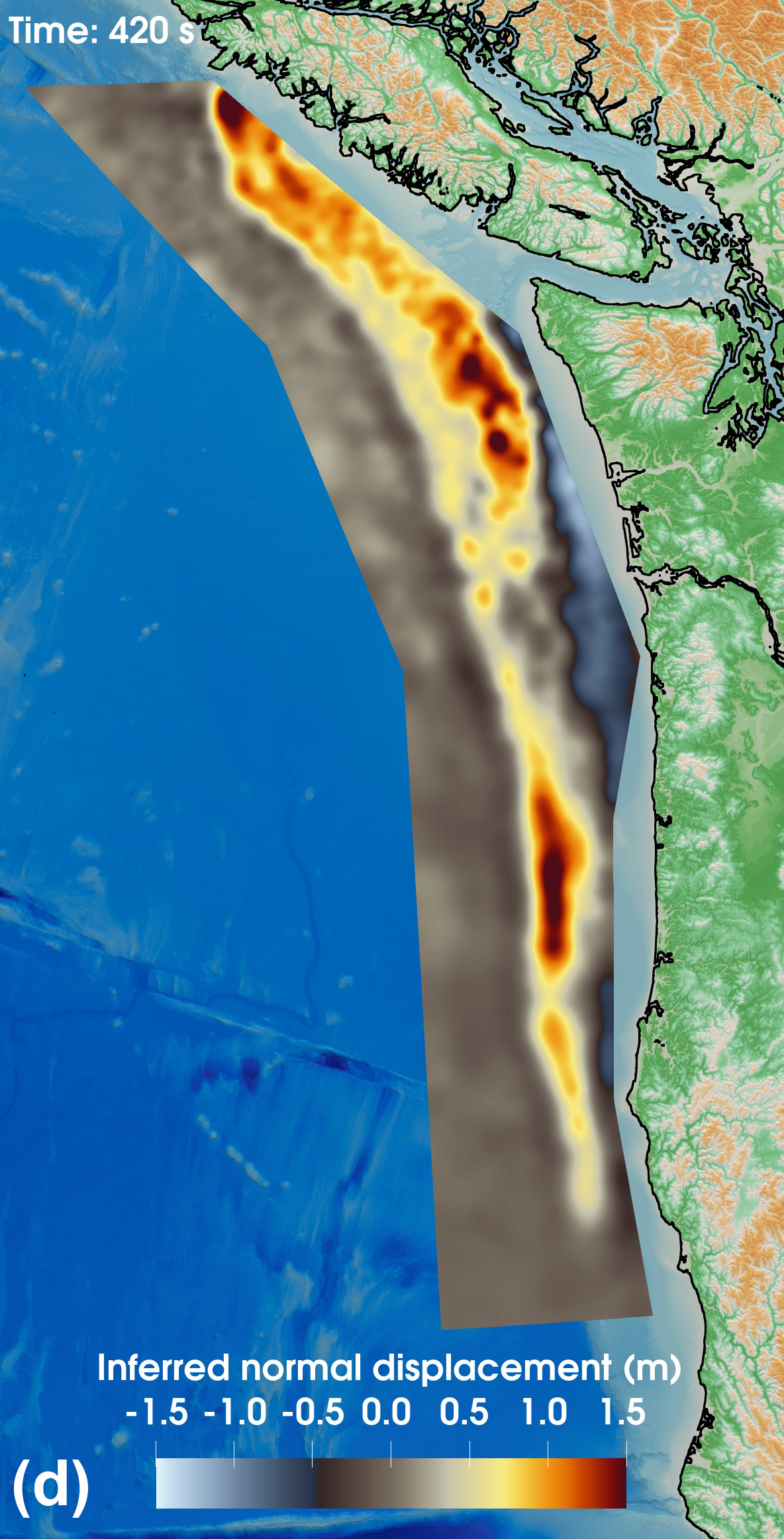}%
    \includegraphics[width=0.166\textwidth]
    {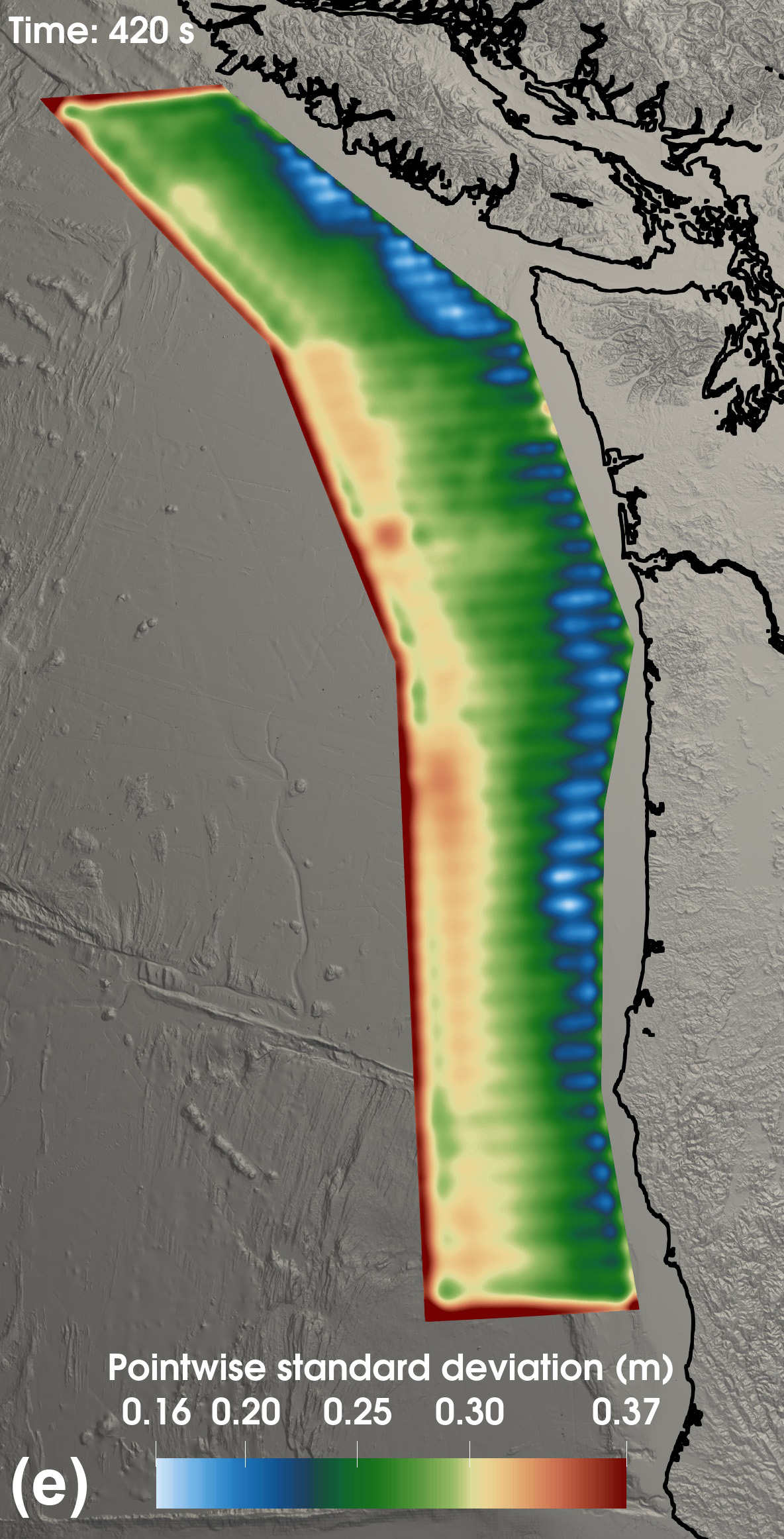}%
    \includegraphics[width=0.166\textwidth]
    {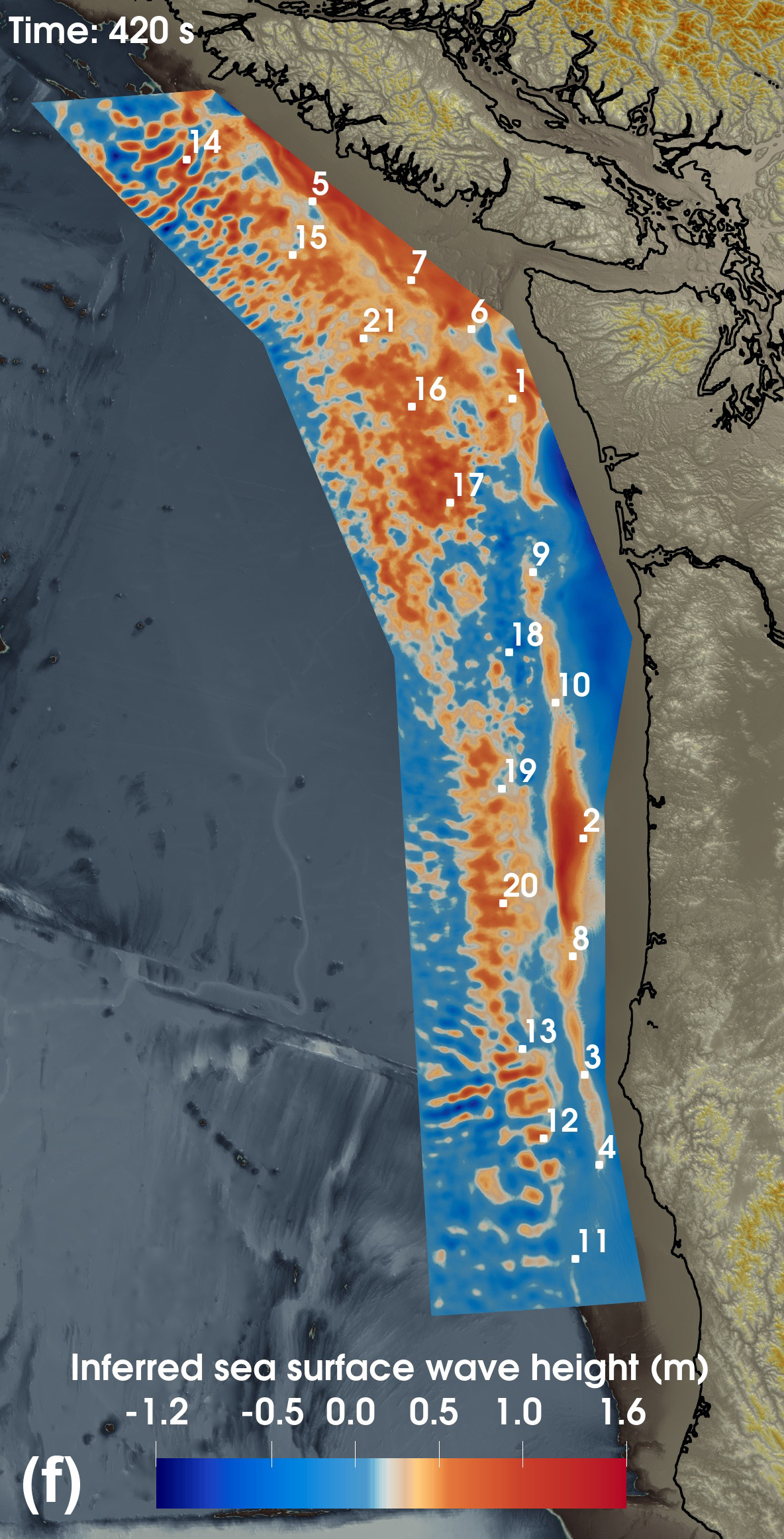}
    \caption{Physics-based magnitude 8.7 dynamic rupture earthquake scenario for a margin-wide rupture in the CSZ, from left to right: 
    (a)~true seafloor displacement; 
    (b)~snapshot of true seafloor acoustic pressure field with 600 hypothesized sensor locations; 
    (c)~snapshot of true sea surface wave height;
    (d)~inferred mean of seafloor displacement; 
    (e)~uncertainties plotted as pointwise standard deviations in meters of seafloor normal displacement;
    and (f)~snapshot of reconstructed sea surface wave height with 21 locations for QoI predictions.
    Hyperlinks to animations of:
    (i)
    \href{https://youtu.be/eQlHehX_u6k}{source}
    (seafloor vertical uplift and normal velocity);
    (ii)
    \href{https://youtu.be/-CPxuK6bebk}{forward solution}
    (seafloor pressure and surface wave height); and
    (iii)
    \href{https://youtu.be/9OAPWumAd1g}{inverse solution}
    (true and inferred seafloor normal displacement).}
    \label{fig:csz-mw}
\end{figure*}

\subsection{FFT-based Hessian matvecs}\label{sec:fast-hessian-matvecs}


Solving the inverse problem \eq{mmap-problem} requires numerous Hessian matvecs, each of which comes at the cost of a pair of forward/adjoint PDE solutions.
For large-scale problems, PDE solutions are computationally expensive and cannot be performed in real time. 
One key innovation that enables real-time inference is to recognize and exploit the fact that the p2o map is governed by an \emph{autonomous dynamical system}.
Specifically, the acoustic--gravity model \eq{fwd-pde} represents a linear time-invariant (LTI) dynamical system for which the PDE operator does not explicitly depend on time.

The LTI dynamical system structure of the governing equations implies that the discrete p2o map $\ptomat$, which maps parameters $\paramvec$ (spatiotemporal seafloor velocity) to observables $\datavec$ (seafloor pressure) via solution of the discretized equations of the acoustic--gravity PDE model \eq{fwd-pde}, inherits a special structure.
Let $\numparam$ denote the spatial parameter points, $\numdata$ the number of sensors ($\numdata \ll \numparam$), and $\numtime$ the temporal dimension of parameters and observations\footnote{For simplicity of presentation, and without loss of generality, we assume that the temporal discretization of the parameters is taken the same as the frequency of the observations and QoI predictions, i.e., $\numtime = \numtime^d = \numtime^q$.} ($\numtime \gg 1$), i.e.,
\begin{itemize}
	\item $\paramvec \in \bb R^{\numparam \numtime}$ with blocks $\paramvec_j \in \bb R^{\numparam}$, $j = 1,2, \ldots, \numtime$;
	\item $\datavec \in \bb R^{\numdata \numtime}$ with blocks $\datavec_i \in \bb R^{\numdata}$, $i = 1,2, \ldots, \numtime$.
\end{itemize}

Then, the discrete p2o map $\ptomat$ can be written as a \emph{block lower-triangular Toeplitz} matrix:
\bes
	\left[ \begin{array}{@{}c@{}}
	\datavec_1 \\[2pt]
	\datavec_2 \\[6pt]
	\datavec_3 \\[1pt]
	\vdots \\[3pt]
	\datavec_{\numtime}
	\end{array} \right]
	=
	\left[ \begin{array}{@{}ccccc@{}}
	\ptomat_{11} & \vb 0 & \vb 0 & \cdots & \vb 0 \\[2pt]
	\ptomat_{21} & \ptomat_{11} & \vb 0 & \cdots & \vb 0 \\
	\ptomat_{31} & \ptomat_{21} & \ptomat_{11} & \ddots & \vdots \\
	\vdots & \vdots & \ddots & \ddots & \vb 0 \\[2pt]
	\ptomat_{\numtime,1} & \ptomat_{\numtime-1,1} & \cdots & \ptomat_{21} & \ptomat_{11}
	\end{array} \right]
	\left[ \begin{array}{@{}c@{}}
	\paramvec_1 \\[2pt]
	\paramvec_2 \\[6pt]
	\paramvec_3 \\[1pt]
	\vdots \\[3pt]
	\paramvec_{\numtime}
	\end{array} \right] ,
	\label{eq:ShiftInvariance}
\ees
where $\ptomat \in \bb R^{(\numdata \numtime) \times (\numparam \numtime)}$ with blocks $\ptomat_{ij} \in \bb{R}^{\numdata \times \numparam}$, $i,j = 1,2, \ldots, \numtime$.


In other words, the block Toeplitz structure of $\ptomat$ corresponds to a \emph{time-shift invariance} of its blocks $\ptomat_{ij}$.
This shift invariance implies that
(i) the p2o map can be precomputed from only $\numdata$ (number of sensors) adjoint PDE solutions (corresponding to the first block column of $\ptomat$) and stored compactly at cost of $\mc{O}(\numparam \numdata \numtime)$ memory;
(ii) the block Toeplitz matrix can be embedded within a block circulant matrix that is block-diagonalized by the discrete Fourier transform \cite{gray2006toeplitz};
and (iii) the p2o matvec then becomes a block-diagonal matvec operation in Fourier space.

These FFT-based p2o matvecs can be implemented efficiently on multi-GPU clusters. In particular, care is taken to arrange the data layouts of the matrices and vectors in each part of the computation to minimize the amount of GPU--GPU communication. By exchanging the order of space and time vector indices local to each processor, the need for strided memory accesses during the core computational kernels is avoided. A 2D GPU partitioning scheme is employed, where the dimensions of the processor grid are adaptively tuned according to the problem sizes and total number of GPUs in order to further reduce communication costs~\cite{venkat2024fft}.

The main advantage of the FFT-based matvec algorithm is that it avoids PDE solves entirely, thus making it independent of the cost of state discretization, time-stepping constraints from Courant--Friedrichs--Lewy (CFL) condition, and computing on unstructured mesh-based data structures dictated by the PDE discretization.
Moreover, it relies on arithmetic operations that involve data that is contiguous in memory and leverages routines from libraries such as cuBLAS that are well known to achieve high performance on GPUs.
As we shall see, the initial cost of precomputing $\ptomat$, for a 
practical number of sensor locations, is easily amortized in the context of solving Bayesian inverse problems, where the matrix-free Hessian actions (that involve costly PDE solutions) can now be replaced with these extremely
fast FFT-based Hessian matvecs.

Finally, the entire discussion about the p2o map $\ptomat$ analogously applies to the p2q map $\ptqmat \in \bb R^{(\numqoi \numtime) \times (\numparam \numtime)}$ that maps the parameters $\paramvec$ to the QoI $\qoivec \in \bb R^{\numqoi \numtime}$ (sea surface wave heights) at $\numqoi$ forecast locations ($\numqoi \ll \numparam$).

\subsection{Real-time Bayesian inference and prediction framework}

To enable real-time Bayesian inference, our framework decomposes the inverse solution into several precomputation (offline) Phases 1--3 and a real-time inference  (online) Phase~4 \cite{henneking2025goal}, as depicted in Fig.~\fig{workflow}.

Phase~1 performs $\numdata + \numqoi$ (number of sensors plus QoI forecasts) adjoint PDE solutions to precompute the block Toeplitz matrices $\ptomat$ and $\ptqmat$.

Phase~2 computes the block Toeplitz matrices $\{ \Gmat^*, \Gmat_q^* \} = \bfpriorcovm \{ \ptomat^*, \ptqmat^* \}$ by performing $\numdata + \numqoi$ solves of the inverse elliptic operator defining the prior covariance $\bfpriorcovm$. Then, the Sherman--Morrison--Woodbury formula is used
to rewrite the posterior covariance as
\bes
	\bfpostcovm
    = \left( \ptomat^* \bfnoise^{-1} \ptomat + \bfpriorcovm^{-1} \right)^{-1} \!
    = \left( \bfpriorcovm - \Gmat^* \Kmat^{-1} \Gmat \right) ,
\ees
where $\Kmat = \bfnoise + \ptomat \Gmat^*$.
Doing so effectively shifts the inverse operator from the intractable high-dimensional parameter space (of dimension $\numparam \numtime$) to the still-large-but-tractable data space (of dimension $\numdata \numtime$); for that reason, we refer to $\Kmat$ as the (prior-preconditioned) ``data space Hessian.''
The dense $\Kmat$ matrix is precomputed with $\numdata \numtime$ matvecs on unit vectors, each of which---thanks to the FFT-based matvecs of $\ptomat$ and $\Gmat^*$---can be executed within fractions of a second, after which $\Kmat$ can be Cholesky-factorized.

Phase~3 computes the uncertainties for the QoI predictions by forming the dense covariance matrix $\bfpostcovq = \ptqmat \bfpostcovm \ptqmat^*$. This requires $\numqoi\numtime$ matvecs with $\bfpostcovq$ on  unit vectors; each matvec effectively solves a billion-parameter inverse problem (in the actions of $\bfpostcovm$ on a vector), which is made tractable thanks to our ability to solve the inverse problem in a fraction of a second.  
Analogously, we precompute the data-to-QoI map, $\Qmat \coloneq \ptqmat \bfpostcovm \ptomat^* \bfnoise^{-1}$, which, for a moderate number of QoI forecasts, can (in Phase~4) be used to make rapid real-time QoI predictions directly from observational data, thereby bypassing the need to 
explicitly reconstruct the seafloor velocity parameters  
(see \S\ref{sec:results}).

Phase~4 executes the (online) parameter inference and QoI prediction, which, given data $\obsvec$, reduces to computing the MAP points, 
$\bfmmap = \bfpostcovm \ptomat^* \bfnoise^{-1} \obsvec$ and $\bfqmap = \ptqmat \bfmmap = \Qmat \obsvec$ 
(taking $\bfmprior = \boldsymbol{0}$),
both of which, after the precomputations done in Phases~1--3, can be executed \emph{in real time}.

\begin{figure}[htb]
    \includegraphics[width=\columnwidth]
    {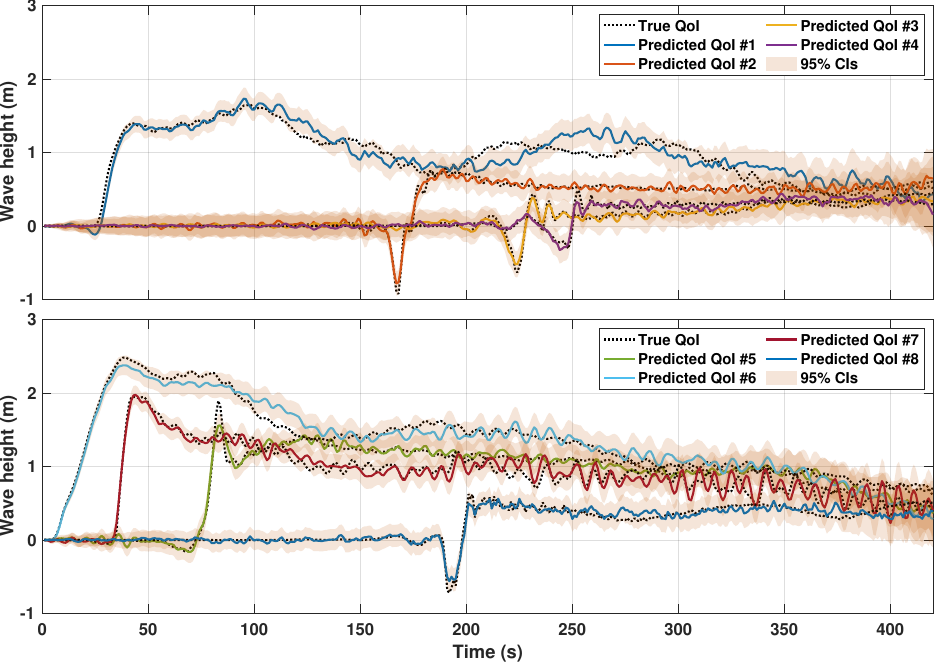}
    \caption{Real-time QoI predictions with uncertainties illustrated as 95\% credible intervals (CIs) inferred from noisy, synthetic data of 600 hypothesized seafloor acoustic pressure sensors for a margin-wide rupture in the CSZ. The QoI numbers (\#1--\#8) refer to (a subset of) the 21 QoI forecast locations marked in the inferred (reconstructed) sea surface wave height plot in Fig.~\ref{fig:csz-mw}.}
    \label{fig:csz-qoi}
\end{figure}

\subsection{Application to the Cascadia subduction zone}

To assess the feasibility of our framework for near-field real-time tsunami forecasting in Cascadia without assuming simplified earthquake models, we generated seafloor displacements that capture the nonlinear interaction of seismic wave propagation and frictional rock failure during a realistic CSZ rupture scenario computed with a 3D dynamic earthquake rupture model \cite{glehman2024partial,uphoff2017sumatra}.
We then used these ``true'' seafloor displacements as a source for our acoustic--gravity forward model (i.e.~normal displacements $b(\vec x, t)$ in \eq{fwd-pde}) to generate synthetic observational data with 1\% relative added noise, at $\numdata=600$ hypothesized seafloor pressure sensors. Using these sparse, noisy observations of acoustic pressure, we then solved the PDE-based inverse problem for the spatiotemporal parameter field representing seafloor velocities during the rupture event, and made predictions of the QoI (sea surface wave height) at $\numqoi=21$ forecast locations.
We note that even a small number of sea surface wave height forecasts is informative for early warning purposes \cite{liu2021comparison}.

To solve this inverse problem, we discretized the acoustic--gravity model \eq{fwd-pde} with high-order finite elements in MFEM (see \S\ref{subsec:MFEM}) using a hexahedral mesh with a spatial resolution of 300~meters (less in shallow areas of the CSZ) with 3.74 billion states and explicit Runge--Kutta time-stepping, with a timestep size dictated by the CFL condition, for a simulation time of 420~seconds.
The parameter field was discretized with $\numparam=$~2,416,530 spatial points and $\numtime=$~420 temporal steps (1~Hz frequency) for a total parameter dimension of $\sim$1.015 billion.
Results of the parameter inference and QoI prediction under uncertainty are shown in Figs.~\fig{csz-mw} and \fig{csz-qoi}; note that the inverse solution is depicted in terms of the seafloor normal displacement, $\int_0^T m(\vec x, t) dt$, which is visually simpler to interpret than snapshots of the inferred spatiotemporal parameter field (seafloor velocities).
The computation times and other performance-related aspects for this inverse problem are presented in \S\ref{sec:results}.

%% file: 6_measurement.tex
Our inference and prediction framework consists of two code bases.
The first is the ``Cascadia application code,'' a C++ 
solver for the acoustic--gravity model \eq{fwd-pde} implemented in MFEM.
This code is used in the (offline) Phase~1 of our framework to compute adjoint PDE solutions that define
the p2o map $\ptomat$ and p2q map $\ptqmat$.
Both AMD (via HIP) and NVIDIA (via CUDA) GPU-based systems are supported; we report performance results for both types of systems.

The second code, implementing Phases~2--4, consists of two parts. The first is ``FFTMatvec,'' which computes matvecs with the block Toeplitz matrices $\ptomat$ and $\ptqmat$ using a portable C++/CUDA/HIP implementation based on~\S\ref{sec:fast-hessian-matvecs}. Performance is reported on AMD and NVIDIA systems. The second part solves the real-time Bayesian parameter inference and QoI prediction problems using the algorithmic innovations in~\S\ref{sec:innovations}. It remains CUDA-exclusive, using libraries like cuDSS and cuSOLVERMp; performance is reported on NVIDIA systems.

All computations are performed in double precision; for ill-posed inverse problems like ours, single precision is unstable.

\subsection{HPC systems}
LLNL's \emph{El Capitan} is an HPE Cray EX supercomputer with 11,136 nodes, each containing 4 AMD Instinct MI300A Accelerated Processing Units (APUs). The MI300A consists of 24 Zen4-based CPU cores, a CDNA3-based GPU, and 128~GB of HBM3 memory, integrated onto a single package. 
Each MI300A has a 61.3~TFLOP/s peak double precision throughput for a total machine peak of 2.73~EFLOP/s.
The system is connected by an HPE Slingshot-200 dragonfly interconnect with at most three network hops between any two nodes.

CSCS's \emph{Alps} is an HPE Cray EX supercomputer with 2,688 nodes, each containing 4 NVIDIA GH200 superchips. 
Each GH200 integrates a Grace CPU (72 Arm cores)
and an H100 GPU (96~GB HBM3, 34.0~TFLOP/s peak double precision throughput); the system peak is 574.8~PFLOP/s.
The system is connected by an HPE Slingshot-11 dragonfly interconnect.

NERSC's \emph{Perlmutter} is an HPE Cray Shasta supercomputer with 1,536 nodes, each containing an AMD EPYC 7763 64-core CPU and 4 NVIDIA A100 GPUs. Each GPU has 40~GB of HBM2 memory and a peak double precision throughput of 9.7~TFLOP/s for a total system peak of 59.6~PFLOP/s. The system is connected by a Slingshot-11 dragonfly interconnect.

\subsection{MFEM library}
\label{subsec:MFEM}
MFEM is a C++ finite element (FE) library that provides high-performance discretization algorithms to a wide range of HPC application codes in national laboratories, industry, and academia \cite{anderson2021mfem}. In addition to SoA support for high-order methods, MFEM has been designed to be highly scalable on the full range of computing platforms, from laptops to GPU-accelerated supercomputers \cite{mfem2024}. Building on efforts in the U.S.\ Exascale Computing Project \cite{ceed}, the library has put a special emphasis on GPU architectures.
An important algorithmic innovation that enables MFEM to achieve high performance on GPU hardware is the concept of \emph{FE operator decomposition} that exposes data parallelism in high-order FE computations \cite{ceed2020scalability}, and enables additional optimizations on quadrilateral and hexahedral grids by taking advantage of the Cartesian product structure at the element level in the decomposition. Critically, MFEM's approach also leads to a significantly reduced memory requirement through \emph{partial assembly} (PA), which stores an asymptotically optimal amount of data: $\mc{O}$(1) per degree of freedom (DOF). Compared to the classical full (global sparse matrix) or element-level (local dense matrices) assembly algorithms, 
PA
leads to orders-of-magnitude faster execution times on both CPU and GPU architectures. Another option supported by MFEM is \emph{matrix-free} (MF) assembly where no data is stored and all computations are done on the fly; see Fig.~\ref{fig:cascadia-optimizations}.
In this work, we focus on faster time-to-solution (as opposed to higher FLOP/s), so we use the PA approach to discretize the forward problem~\eqref{eq:fwd-pde}. 

\subsection{Cascadia application code}

The acoustic--gravity forward problem \eq{fwd-pde}, as well as its adjoint problem, were cast into a mixed variational formulation and then discretized with MFEM using the Galerkin FE method and explicit fourth-order Runge--Kutta (RK4) time-stepping. The FE discretization uses fourth-order continuous ($H^1$-conforming) scalar-valued pressure and third-order discontinuous ($L^2$-conforming) velocity components.

\begin{table}[htb]
\centering
\caption{Cascadia application code: timers}
\label{tab:cascadia-timers}
\begin{tabular}{ll}
    \toprule
    Timer & Main tasks \\
    \midrule
    \rowcolor{GREY!5}
    Initialization
    & Initialize MPI devices \\
    Setup
    & Read and partition mesh \\
    & Partially assemble the mixed FE operator \\
    & Precompute parameter-to-state mapping \\ 
    & Precompute state-to-observation operator \\
    \rowcolor{GREY!5}
    Adjoint p2o/p2q
    & Apply transpose observation operator (RHS) \\
    \rowcolor{GREY!5}
    & Solve adjoint acoustic--gravity wave propagation \\
    \rowcolor{GREY!5}
    & Retrieve p2o/p2q column vector from adjoint state \\
    I/O
    & Write adjoint p2o/p2q column vector to file \\
    \bottomrule
\end{tabular}
\end{table}

Table~\ref{tab:cascadia-timers} summarizes the main steps of the Cascadia application for computing adjoint PDE solutions in Phase~1 of our framework.
The timers represent wall time; they are implemented with standard POSIX clock functions, called after \texttt{\{hip,cuda\}DeviceSynchronize} and \texttt{MPI\_Barrier}.

While performance is measured for the entire application, the computational expense is overwhelmingly dominated by the cost of the acoustic--gravity wave propagation solver.
In particular, the dominant cost is the repetitive application (four applications per RK4 timestep) of the time-stepping operator
\bes
    \left[
    \vb u_\delta \ \vb p_\delta
    \right]^T
    = \vb M^{-1}
    \left(
    - \vb A
    \left[
    \vb u \ \vb p
    \right]_i^T
    +
    \left[
    \vb f \ \vb g
    \right]_i^T
    \right) ,
\ees
where $\left[ \vb u \ \vb p \right]_i^T$ and $\left[ \vb f \ \vb g \right]_i^T$ are respectively the state and right-hand-side (RHS) vectors at time instance $i$, and $\left[ \vb u_\delta \ \vb p_\delta \right]$ is the state increment used by RK4. The (lumped) mass matrix $\vb M$ and stiffness matrix $\vb A$ are discretizations of the block operators $M$ and $A$ defined by:
\bes
    \! \! \! \! \left( \!
    M \!
    \left[
    \begin{array}{@{}c@{}}
        \vec u \\
        p
    \end{array}
    \right] \! , \!
    \left[
    \begin{array}{@{}c@{}}
        \vec \tau \\
        v
    \end{array}
    \right]
    \right)
    \coloneq
    \left[
    \begin{array}{@{}cc@{}}
        (\rho \vec u, \vec \tau) & 0 \\
        0 & \! \! \! (K^{-1} p, v) + \lb(\rho g)^{-1} p, v \rb_{\p \Omega_{\text s}} \\
    \end{array}
    \right]
    \! \!
\ees
and
\be
    \left( \!
    A \!
    \left[
    \begin{array}{@{}c@{}}
        \vec u \\
        p
    \end{array}
    \right] \! , \! 
    \left[
    \begin{array}{@{}c@{}}
        \vec \tau \\
        v
    \end{array}
    \right]
    \right)
    \coloneq
    \left[
    \begin{array}{@{}cc@{}}
        0 & (\nabla p, \vec \tau) \\
        -(\vec u, \nabla v) & \lb Z^{-1} p, v \rb_{\p \Omega_{\text a}} \\
    \end{array}
    \right] ,
    \label{eq:block-mult}
\ee
where $\vec u, \vec \tau \in (L^2(\Omega))^3$ and $p, v \in H^1(\Omega)$; $(\cdot, \cdot)$ denotes the (component-wise) $L^2(\Omega)$ inner product, and $\lb \cdot, \cdot \rb_{\p \Omega}$ is the $L^2(\p \Omega)$ inner product over (part of) the boundary $\p \Omega$.

\begin{table}[htb]
\centering
\caption{Scalability setup on \tc{BLUE}{El Capitan}, \tc{OLIVE}{Alps} and \tc{RED}{Perlmutter}}
\label{tab:elcap-scaling-setup}
\begin{tabular}{rrrrr}
    \toprule
    & & & Weak scaling & Strong scaling \\
    \cmidrule{4-5}
    Nodes & GPUs & Processor grid & Mesh elements & Elements/GPU \\
    \midrule
    \rowcolor{BLUE!25}
    85 & 340 & 5 $\times$ \hspace{1pt} 17 $\times$ 4 & 1,693,450,240 & 4,980,736 \\
    \rowcolor{BLUE!25}
    10,880 & 43,520 & 80 $\times$ 136 $\times$ 4 & 216,761,630,720 & 38,912 \\
    \midrule
    \rowcolor{OLIVE!25}
    36 & 144 & 2 $\times$ \hspace{1pt} 18 $\times$ 4 & 566,231,040 & 3,932,160 \\
    \rowcolor{OLIVE!25}
    2,304 & 9,216 & 16 $\times$ 144 $\times$ 4 & 36,238,786,560 & 61,440 \\
    \midrule
    \rowcolor{RED!25}
    47 & 188 & 1 $\times$ \hspace{1pt} 47 $\times$ 4 & 295,698,432 & 1,572,864 \\
    \rowcolor{RED!25}
    1,504 & 6,016 & 8 $\times$ 188 $\times$ 4 & 9,462,349,824 & 49,152 \\
    \bottomrule
\end{tabular}
\end{table}

The primary performance metric that matters from our application point of view is therefore \emph{runtime per timestep}, which we report, together with parallel efficiency and relative speedup, in our weak and strong scalability studies.
Reported runtimes were averaged over ten consecutive timesteps (40 operator applications), discarding the initial ten timesteps as warm-up.
The mesh sizes we used to test scalability on \emph{El Capitan}, \emph{Alps}, and \emph{Perlmutter} are summarized in Table~\tab{elcap-scaling-setup}.
Since runtime per timestep is the primary application-relevant performance metric, 
we optimized GPU-kernel performance
for \emph{DOF throughput} rather than total achieved FLOP/s (see \S\ref{subsec:peak_performance}); 
FLOP and byte counts were manually calculated, and memory access rates were analyzed using the ROCm profiler on \emph{El Capitan} and Nsight Compute on \emph{Alps}.


\subsection{FFT-based Bayesian inference code}
\label{subsec:fft-code}
The performance-portable, open-source FFTMatvec\footnote{\url{https://github.com/s769/FFTMatvec}} is application agnostic and can be used with any block-triangular Toeplitz matrix; see~\S\ref{sec:implications}. It uses vendor libraries (e.g., \{cu,roc\}FFT, \{cu,roc\}BLAS) for batched computations. Its performance was profiled with Nsight Compute and rocprofv3.

The CUDA-based Bayesian inference code uses cuDSS for prior solves to form the $\vb{G}^*$ matrix and cuSOLVERMp for linear solves with the dense, symmetric $\vb{K}$ matrix via Cholesky factorization. Wall timings for FFTMatvec and the Bayesian inference codes were measured with \texttt{omp\_get\_wtime} after device synchronization and \texttt{MPI\_Barrier}. The primary performance goal of both codes is a fast time-to-solution.

%% file: 7_results.tex
\subsection{Scalability}

\begin{figure*}[htb]
    \includegraphics[width=0.49\textwidth]
    {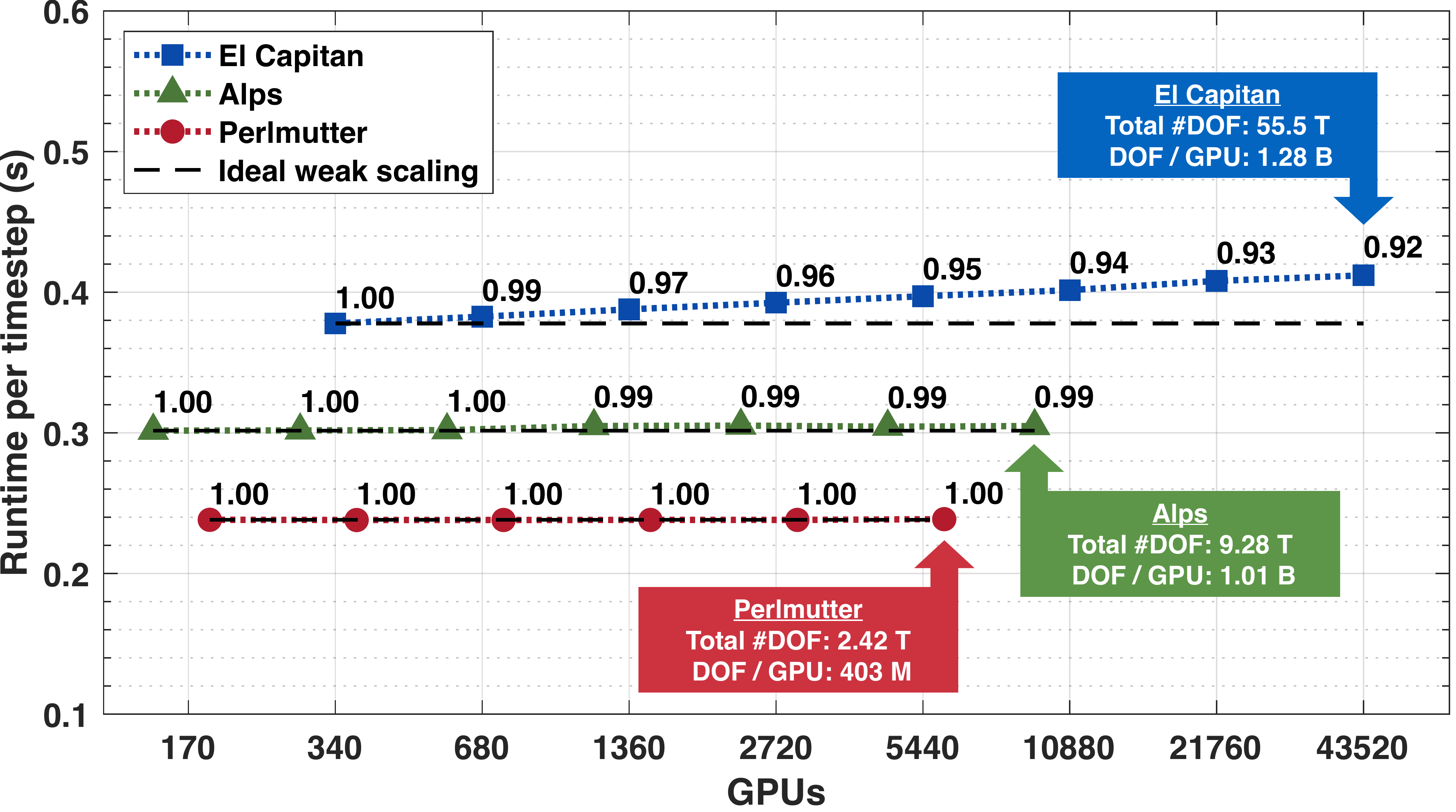}
    \hspace{0.01\textwidth}
    \includegraphics[width=0.49\textwidth]
    {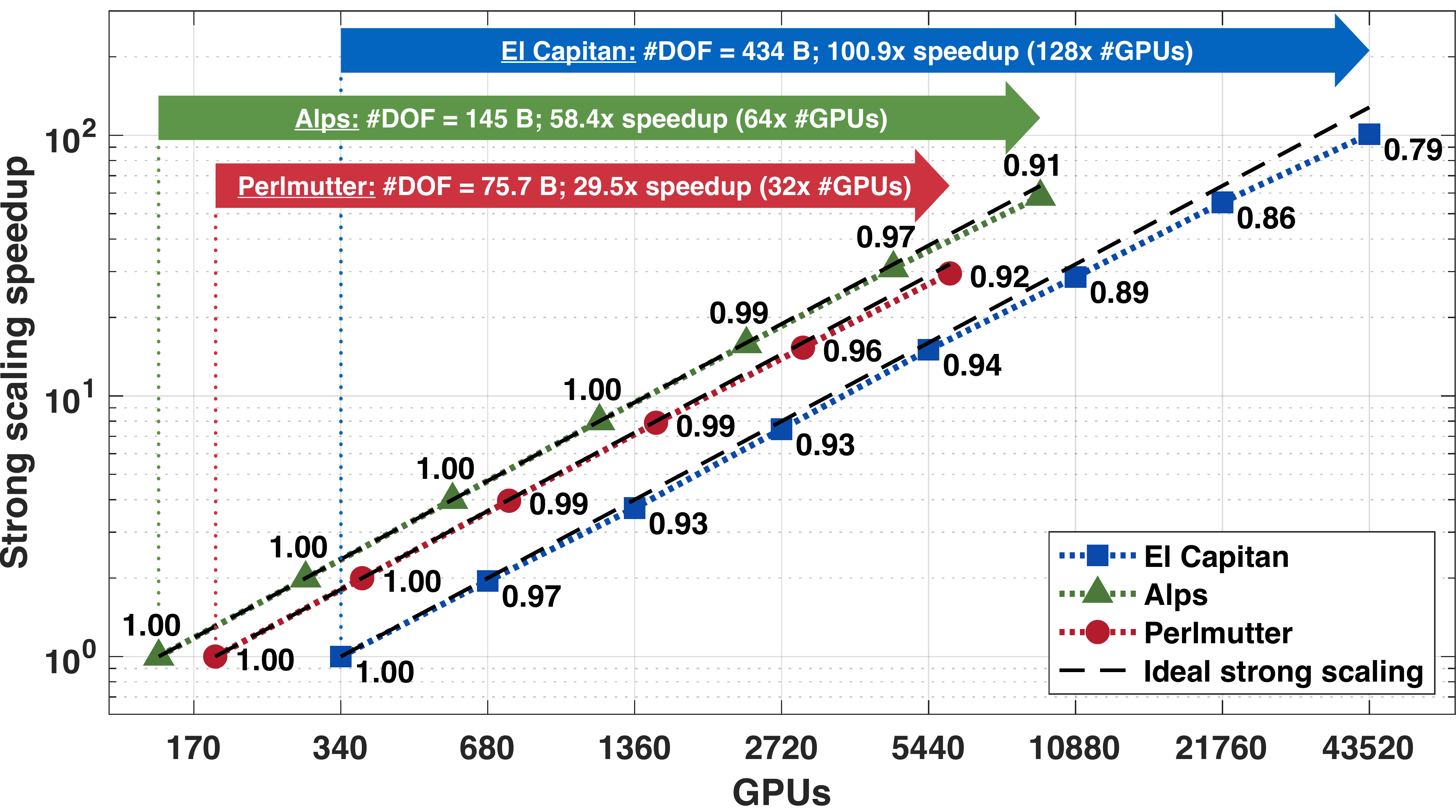}
    \caption{Weak scalability (left) and strong scalability (right) results on \emph{El Capitan}, from 85 nodes (340 AMD MI300A GPUs) to 10,880 nodes (43,520 GPUs), on \emph{Alps}, from 36 nodes (144 NVIDIA GH200 GPUs) to 2,304 nodes (9,216 GPUs), and on \emph{Perlmutter}, from 47 nodes (188 NVIDIA A100 GPUs) to 1,504 nodes (6,016 GPUs). Numbers along the graph lines indicate parallel efficiency.}
    \label{fig:scaling}
    \vspace{-7pt}
\end{figure*}

We present weak and strong scalability results (Fig.~\ref{fig:scaling}) on the \emph{El Capitan}, \emph{Alps}, and \emph{Perlmutter} supercomputers.
The computational cost of our inference and prediction framework is overwhelmingly dominated by the cost of the adjoint wave propagation PDE solutions in Phase~1 (offline) that are used to construct the p2o map $\ptomat$ and p2q map $\ptqmat$.
The computational cost of each adjoint PDE solution, in turn, is overwhelmingly dominated by the RK4 time-stepping solver; each timestep involves four applications of the PDE operator. Scalability to large problem sizes is critical (in addition to throughput) for this offline phase, since we wish to apply our framework to other settings beyond Cascadia where tsunamis can impact populations across long distances (such as occurred in the 2004 Sumatra--Andaman earthquake).

On \emph{El Capitan}, the solver maintained 92\% weak parallel efficiency over a 128-fold increase in problem size, from 85 nodes (340 AMD MI300A GPUs) to 10,880 nodes (43,520 GPUs).
The largest problem involved over 55.5 trillion DOF, which, to the best of our knowledge, is \emph{the largest reported unstructured mesh FE computation}.
This computation used 86\% (4.8~PB) of the available machine memory (5.57~PB).
For the largest problem fitting on 340 GPUs (434 billion DOF), the solver achieved a parallel speedup of 100.9 over a 128-fold increase in GPUs---a strong parallel efficiency of 79\%. 

We also obtained excellent weak and strong scalability on \emph{Alps} over a 64-fold increase in GPUs, from 36 nodes (144 NVIDIA GH200 GPUs) to 2,304 nodes (9,216 GPUs). 
Here, the solver achieved 99\% weak parallel efficiency with 1.01 billion DOF per GPU and 91\% strong parallel efficiency for a problem involving 145 billion DOF.
Similar scalability was observed on \emph{Perlmutter} over a 32-fold increase in GPUs.

While GPU-based systems have been the focus of our performance optimizations, we want to emphasize that our solver also exhibits excellent scalability on CPU-based systems.
On TACC's \emph{Frontera}\footnote{\emph{Frontera} hosts 8,368 Intel Xeon Platinum 8280 (``Cascade Lake'') compute nodes contained in 101 racks. Each node has 56 cores on two sockets and 192~GB of DDR4 RAM; see \url{https://docs.tacc.utexas.edu/hpc/frontera/\#system}.} supercomputer, we achieve an outstanding weak parallel efficiency of 95\% over an 8,192-fold increase in problem size, from 1 node (56 cores) to 8,192 nodes (458,752 cores), with the largest problem involving 2.20 trillion DOF (4.80 million DOF per core), and a strong parallel efficiency of 70\% over a 128-fold increase in cores, from 3,584 cores to 458,752 cores.

\begin{figure}[htb]
    \includegraphics[width=0.49\columnwidth]
    {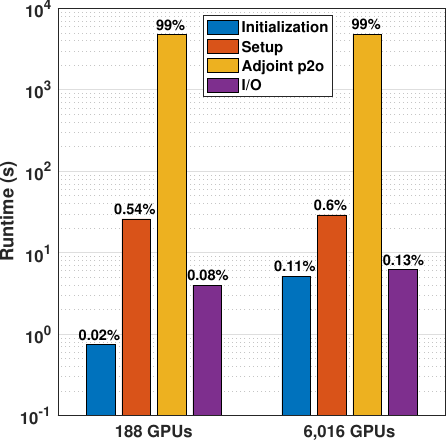}
    \includegraphics[width=0.49\columnwidth]
    {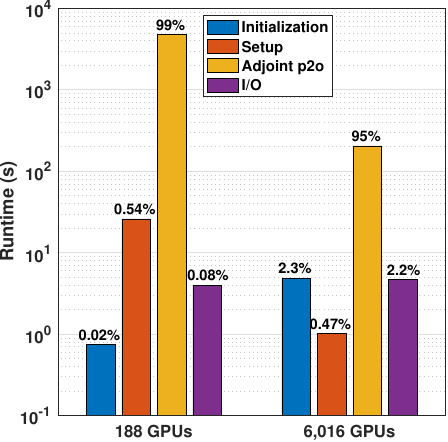}
    \caption{Application timers (as defined in Table~\ref{tab:cascadia-timers}) in the weak scalability (left) and strong scalability (right) limit on \emph{Perlmutter}. Adjoint p2o and I/O timers, each measured for 200 timesteps, are projected to 20,000 timesteps. Numbers on bars indicate percentage of application runtime. Note log scale.}
    \label{fig:cascadia-timers}
\end{figure}

Since small timesteps are needed to resolve fast acoustic waves, the total number of timesteps for the wave propagation solver is usually large. For example, computation of the CSZ rupture scenario on our FE mesh requires on the order of at
least $\mc{O}(10^4)$ timesteps. Fig.~\ref{fig:cascadia-timers} plots application timers (as defined in Table~\ref{tab:cascadia-timers}), where adjoint p2o and I/O timers, each measured for 200 timesteps, are projected to 20,000 timesteps, to illustrate that the time for job initialization, setup, and I/O is negligible in the weak scalability limit but has minor impact in the strong scalability limit, where the entire application runtime is still overwhelmingly dominated by the solver.

\subsection{Peak performance}
\label{subsec:peak_performance}

\begin{figure}[htb]
    \centering
    \includegraphics
    [width=\columnwidth]
    {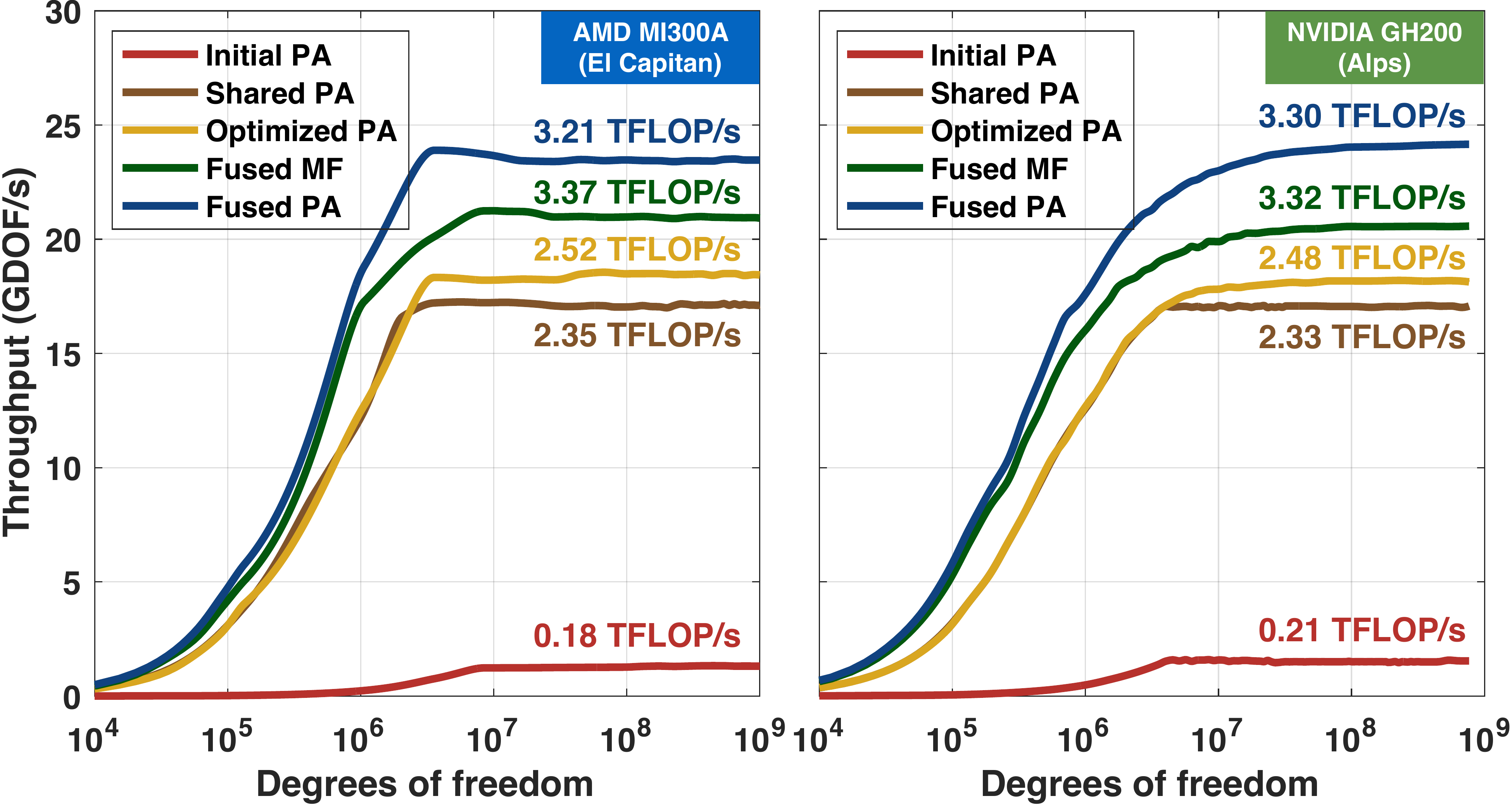}
    \caption{Throughput, in billion degrees of freedom (GDOF) per second, for the kernels corresponding to the off-diagonal blocks in \eqref{eq:block-mult} on a single GPU of \emph{El Capitan} (left) and \emph{Alps} (right). 
    Measured FLOP/s rates 
    are shown in the saturated regime for each curve.
    By utilizing the GPU shared memory we improved the initial partial assembly (PA) kernels (red) to a version (brown) that is 13$\times$ faster. 
    Specifying explicit launch bounds for the kernels allowed us to additionally improve performance (yellow), which was the version of the kernels used in the scaling runs in Fig.~\ref{fig:scaling}. 
    Further optimizations by fusing the actions of the two operators into one kernel allowed us to achieve a peak performance of 24~GDOF/s (blue) corresponding to 3.2~TFLOP/s.
    Although the matrix-free (MF) approach (green) attains higher FLOP/s, its time-to-solution (DOF throughput) remains lower than that of Fused PA.
    The results on \emph{El Capitan} and \emph{Alps} show that our implementation is performance portable.
    }
    \label{fig:cascadia-optimizations}
\end{figure}

While the existing methods in MFEM already scaled well for the forward problem, we introduced additional GPU performance and memory optimizations to further reduce the runtime and increase the size of the problems we can simulate. Some of the performance optimizations for the two key kernels in \eqref{eq:block-mult} are shown in Fig.~\ref{fig:cascadia-optimizations}. These kernels account for the majority of the time in each stage of the RK4 time-stepping.

We also performed extensive optimization of the memory usage, including: instrumenting the code to track memory usage separately on host and device;
taking advantage of the unified memory on the MI300A APU by freeing host allocations so they can be used for GPU allocations; using sparsity in the RHS to avoid storing full vectors; fusing permutation operators with operator action kernels; recomputing determinants of Jacobian matrices instead of storing them; batching multiple small memory allocations together; 
and carefully reusing temporary vectors from RK4 for temporary vectors needed in the operator action. With all these optimizations for a problem of size 67 million DOF per APU, we improved unified memory usage from (5.2 host + 30.7 device) to (1.1 host + 5.64 device) GiB/APU, a reduction of 5.33$\times$. These improvements allowed us to fit $\sim$1.28 billion DOF per APU on \emph{El Capitan} and $\sim$1.01 billion DOF per GPU on \emph{Alps}.

Our primary performance metric is time-to-solution, which, for the PDE operator \eqref{eq:block-mult}, 
is measured as DOF throughput. For a given problem size, higher throughput is directly proportional to faster computation time.
As shown in Fig.~\ref{fig:cascadia-optimizations}, higher FLOP/s does not necessarily lead to faster time-to-solution. 
On MI300A nodes of \emph{El Capitan}, the best-performing implementation, Fused PA, achieves a lower percentage (5.2\%) of theoretical peak FLOP/s than Fused MF (5.5\%) but is faster. 
While Fused MF has a higher arithmetic intensity (7.3~FLOP/byte) than Fused PA (2.4~FLOP/byte), it transfers less data per DOF (22.2 byte/DOF vs.\ 57.0 byte/DOF). Thus, its 1.18$\times$ higher FLOP/DOF ratio offsets the 1.05$\times$ FLOP/s improvement, yielding a 1.12$\times$ longer runtime.
Although the percentage of peak FLOP/s is not as high as for dense matrix operations, we note that it is much better than the 0.63\% of peak reported for the HPCG benchmark\footnote{\url{https://top500.org/lists/hpcg/2025/06}} on \emph{El Capitan}.
Similar to the type of computations performed in FE algorithms, HPCG has a FLOP count and memory access rate proportional to the number of DOF.


All GPU kernels in FFTMatvec (\S\ref{subsec:fft-code}) are memory-bound, achieving 80--95\% of peak memory bandwidth on NVIDIA (A100, GH200) and AMD (MI250X, MI300X) GPUs \cite{venkat2024fft}.


\subsection{Time-to-solution}

We measured time-to-solution for a margin-wide rupture scenario in the CSZ for which we presented inversion results in \S\ref{sec:innovations}. 
These computations were performed on 128 {\em Perlmutter} nodes (512 NVIDIA A100 GPUs) and involved inference of more than one billion parameters from 252,000 synthetic, noisy observations obtained by 600 hypothesized seafloor pressure sensors, followed by real-time forecasting of the QoI (surface wave heights) under uncertainty for 21 QoI locations.

For this problem, Table~\tab{time-to-solution} summarizes the compute times for each phase of our inference and prediction framework. 
As discussed above, the total computational expense is dominated by the PDE solutions in the (offline) Phase~1; scalability is the most important metric in these offline computations, whereas time-to-solution is the metric that matters most for the Phase~4 (online) computations that are executed in real time.

\begin{table}[htb]
    \centering
    \caption{Compute time for each phase of the inference and prediction performed on Perlmutter A100 GPU nodes. Time-to-solution for the \tc{OLIVE}{\textbf{online computation}} (Phase~4) is less than 0.2~seconds.
    }
    \label{tab:time-to-solution}
    \addtolength{\tabcolsep}{-1pt}
\begin{tabular}{rl@{\hspace{5pt}}rl}
    \toprule
    Phase & Task & GPUs & Compute time \\
    \midrule
    \rowcolor{GREY!5}
    1 &form $\ptomat: \paramvec \mapsto \datavec$ & 512&
    600 $\!\times\!$ 52~m $\sim$ 520~h \\
    \rowcolor{GREY!5}
    & form $\ptqmat: \paramvec \mapsto \qoivec$ &512&
    21 $\!\times\!$ 52~m $\sim$ 18~h \\
    \rowcolor{GREY!10}
    2  &form $\Gmat^* \coloneq \bfpriorcov \ptomat^*$ &16&
    600 $\!\times\!$ 4.5~s $\sim$ 45~m \\
    \rowcolor{GREY!10}
    & form $\Gmat_q^* \coloneq \bfpriorcov \ptqmat^*$ &16&
    21 $\!\times\!$ 4.5~s  $\sim$ 1.5~m \\
    \rowcolor{GREY!10}
    &form $\Kmat \coloneq \bfnoise + \ptomat \Gmat^*$ &512&
    252k $\!\times\!$ 24~ms  $\sim$ 100~m \\
    \rowcolor{GREY!10}
    & factorize $\Kmat$ &25& 22~s \\
    \rowcolor{GREY!15}
    3 &compute $\bfpostcovq$ &512& 8,820 $\!\times\!$ 150~ms $\sim$ 25~m \\
    \rowcolor{GREY!15}
    & compute $\Qmat : \datavec \mapsto \qoivec$ &512& 8,820 $\!\times\!$ 150~ms $\sim$ 25~m \\
    \rowcolor{OLIVE!30}
    4 & infer parameters $\bfmmap$ &512& $<$ 0.2~s \\
    \rowcolor{OLIVE!30}
    & predict QoI $\bfqmap$ &1& $<$ 1~ms \\
    \bottomrule
\end{tabular}
\end{table}

Phase~1 relies on the PDE solver to precompute the p2o map $\ptomat$ and p2q map $\ptqmat$.
For the given configuration (600 sensors, 21 QoI locations), we performed a total 621 PDE solutions, each averaging 52~minutes on 512 A100 GPUs, for a total of 538~hours of compute time (note that each one of the 621 PDE solutions can be computed independently).

Once Phase~1 was completed, we exploited the block Toeplitz structure of the precomputed $\ptomat$ and $\ptqmat$ to perform fast FFT-based Hessian matvecs.
Each Hessian matvec, conventionally requiring a pair of forward/adjoint PDE solutions that take 104~minutes on 512 A100 GPUs, can be performed in 0.024 seconds 
on the same number of GPUs---a speedup of 260,000.
We re-emphasize that these FFT-based Hessian matvecs are exact and do not incorporate any approximations. 
The 260,000$\times$ speedup derives primarily from the parallel algorithmic innovations, discussed in \S\ref{sec:innovations}, that massively reduce the overall complexity of the Hessian matvec which, after the precomputations in Phase~1, avoids PDE solutions altogether.

These rapid Hessian matvecs are the workhorse for the remaining offline computations (Phases~2--3).
For example, forming the data space Hessian $\Kmat$ requires 252,000 matvecs, an intractable task if each matvec required a forward/adjoint pair of PDE solutions. Instead, it was carried out using our FFT-based matvecs in just 100~minutes.
Finally, real-time inference of the parameters 
(seafloor motion) 
and forward prediction of the QoI 
(surface wave heights) 
was achieved in \emph{less than 0.2 seconds} of compute time on 512 A100 GPUs. 

We conclude by re-emphasizing that computing the 
inverse solution with SoA methods (see \S\ref{sec:state_of_the_art}), without the innovations described in \S\ref{sec:innovations}, would not only have been prohibitive in real time, but would have been intractable (50 years on 512 A100 GPUs).
%
With the advances described in this paper, the computing time reduces to a one-time offline computation of 621 ($N_d + N_q$) adjoint wave propagations, requiring 538 hours on 512 GPUs and representing an $\sim$810$\times$ reduction in PDE solutions. Then, for each earthquake that excites the sensors, the online solution of the Bayesian inverse problem requires $<0.2$~s, a factor of $10^{10}$ speedup over the SoA CG algorithm. 


%% file: 8_implications.tex

We enable real-time probabilistic tsunami forecasts for near-field events such as a future magnitude 8--9 CSZ megathrust earthquake, where conventional systems may fail due to delayed or inaccurate source characterization. Our performance-portable digital twin framework resolves complex rupture dynamics and tsunami wavefields at unprecedented speed and accuracy, enabling early predictions of inundation and evacuation zones within seconds, critical for life-saving response where warning time is otherwise measured in minutes.
Notably, the real-time inversion itself requires only moderate computational resources. Moreover, if only surface wave heights at selected locations are of interest, the forecasting step reduces to a precomputed, small, dense matrix--vector product---enabling deployment entirely without any HPC infrastructure.

While we demonstrate the potential for real-time forecasting, we remain limited by the sparsity of offshore sensors currently available in the CSZ; however, emerging technologies such as distributed acoustic sensing will improve observational coverage for resolving near-field tsunami source characteristics \cite{Schmidt2023,becerril2024towards}. 
Furthermore, 
expanding to fully-coupled acoustic--elastic simulations \cite{krenz2021palu} allows 
us to employ our framework to invert for fault slip, and forward propagate seismic waves to compute---in real time---maps of the intensity of ground motion in populated regions. These {\em shake maps} provide critical information for early responders and post-earthquake recovery. 

More generally, autonomous dynamical systems arise in many different settings beyond geophysical inversion. Our Bayesian inversion-based digital twin framework is thus more broadly applicable to 
acoustic, electromagnetic, and elastic inverse scattering; source inversion for transport of atmospheric or subsurface hazardous agents; satellite inference of emissions; and treaty verification, among others. 